\documentclass{aa}
\usepackage{graphicx}
\usepackage[varg]{txfonts}
\usepackage{longtable}
\usepackage{lscape}

\begin{document}

 \title{Variability of stellar granulation and convective blueshift with spectral type and magnetic activity.\thanks{Tables A.1 and B.1 are only available in electronic form at the CDS via anonymous ftp to cdsarc.u-strasbg.fr (130.79.128.5) or via http://cdsarc.u-strasbg.fr/viz-bin/qcat?J/A+A/} I. K and G main sequence stars}

   \titlerunning{Variability of stellar granulation and convective blueshift. I. }

   \author{N. Meunier \inst{1}, A.-M. Lagrange \inst{1}, L. Mbemba Kabuiku \inst{1}, M. Alex  \inst{1}, L. Mignon \inst{1}, S. Borgniet \inst{1} 
          }
   \authorrunning{Meunier et al.}

   \institute{
Univ. Grenoble Alpes, CNRS, IPAG, F-38000 Grenoble, France\\
  \email{nadege.meunier@univ-grenoble-alpes.fr}
             }

\offprints{N. Meunier}

   \date{Received 3 June 2016 ; Accepted 21 September 2016}

\abstract{In solar-type stars, the attenuation of convective blueshift by stellar magnetic activity dominates the $RV$ (radial velocity) 
variations over the low amplitude signal induced by low mass planets. Models of stars that differ from the Sun will require a good knowledge of the attenuation of the convective blueshift to estimate its impact on the variations. }
{It is therefore crucial to precisely determine not only the amplitude of the convective blueshift for different types of stars, 
but also the dependence of this convective blueshift on magnetic activity, as these are key factors in our model producing 
the $RV$.}
{We studied a sample of main sequence stars with spectral types from G0 to K2 and focused on their temporally averaged properties: the activity level
and a criterion allowing to characterise the amplitude of the convective blueshift. This criterion is derived from the dependence of the 
convective blueshift 
with the intensity at the bottom of a large set of selected spectral lines.
}
{
We find the differential velocity shifts of spectral lines due to convection to depend on the spectral type, the wavelength (this dependence is correlated with the Teff and activity level), and on the activity level. This allows us to quantify the dependence of granulation  properties
on magnetic activity for stars other than the Sun. 
We are indeed able to derive a significant dependence of the convective blueshift on activity level for all types of stars. 
The attenuation factor of the convective blueshift appears to be constant over the considered range of spectral types. 
We derive a convective blueshift which decreases towards lower temperatures, with a trend in close agreement with models for Teff lower than 5800 K, but with a significantly larger global amplitude. Differences also remain to be examined in detail for larger Teff. 
We finally compare the observed $RV$ variation amplitudes with those that could be derived from our convective blueshift
using a simple law and find a general agreement on the amplitude. We also  show that inclination (viewing angle relative to the stellar equator) plays a major role in the dispersion in $RV$ amplitudes. }
{Our results are consistent with previous results and provide, for the first time, an estimation of the convective 
blueshift as a function of Teff, magnetic activity, and wavelength, over a large sample of G and K main sequence stars. }


\maketitle

\section{Introduction}

The understanding of stellar activity and its impact on convection is an important challenge when 
studying the impact of stellar activity on exoplanet detectability because the variability  of the convection attenuation impacts stellar radial velocities ($RV$s). 
In the solar case,  
the inhibition of convection by magnetic fields indeed dominates the stellar signal \cite[][]{meunier10a}. This model has been
confirmed with several observations: the reconstruction of the solar $RV$ using Michelson Doppler Imager (MDI)  
Dopplergrams 
\cite[][]{meunier10}; more recently, $RV$ variations of similar amplitude have been obtained using 
direct observations of the Sun \cite[][]{dumusque15} as well as using indirect solar $RV$ measurements 
from various bodies in solar system observations (Moon, asteroids, Jupiter satellites) by \cite{lanza16} and \cite{haywood16}.

In simulations made by \cite{meunier10a} based on observed solar features 
and later by \cite{borgniet15} based on simulated solar features, it was assumed that the solar convective blueshift 
was attenuated  by a factor of two thirds, on average, in magnetic regions such as plages and the solar network, based on
solar observations by \cite{BS90}. How stellar granulation changes with magnetic field for different 
stellar types is yet unknown. This is, however, a critical factor in estimating the impact of stellar type on the activity-induced $RV$ variations,  as already pointed out by \cite{dravins90}, for example  by extrapolating the work of \cite{borgniet15} to stars other than the Sun.

The determination of the amplitude of the absolute convective blueshift depends on two parameters, for a given coverage by magnetic features: 
1/ How does the convective blueshift depend on spectral type? 2/ Is the attenuation factor, due to magnetic fields, similar  for all stars?

Concerning the first issue, a number of results in the literature already give some clues. 
\cite{dravins81} have shown 
that the dependancy of the velocity derived from each spectral line as a function of their depth is directly related to the properties of solar convection.
Lines
of different depths (intensities) form at different depths in the atmosphere: they are therefore probing regions with different intensity-velocity correlations inside granules, so that their sensitivity differs for different convective blueshifts. 
Two types of measurements reflect this process: 1/ the absolute shifts of spectral line bisectors; 2/ the differential velocity shifts of spectral lines (namely between lines of different depths). The first ones are difficult to measure because of other effects not easy to quantify at the required precision (except for the Sun). The latter seems to have a universal shape according to \cite{gray09}.  
\cite{hamilton99} also show that the bottom of the line position is less sensitive to resolution effects compared to the bisector analysis, which is also interesting when studying a large sample of stars with different $v$sin$i$, making it a powerful  method to estimate these convection properties, something not possible by $RV$ jitter analysis. 

This property has been studied for stars other than the Sun \cite[e.g.][]{gray82,dravins87b,dravins99,hamilton99,landstreet07, 
allendeprieto02,gray09} but always for a very small number of main sequence stars. The analysis has usually been performed using a very small number of lines, \cite{gray09}, or low signal to noise ratio (S/N) spectra. How the measurements are impacted by the choices of selected lines has not been investigated so far. These studies however show that 
velocity flows inside granules and resulting convective blueshifts increase with Teff.
\cite{pasquini11} have studied a larger sample but found a very large dispersion of the convective blueshift of the main sequence stars.  
On the other hand, recent hydrodynamical numerical simulations have been performed by \cite{allendeprieto13} and show a clear dependence 
of the convective blueshift on spectral type. 
Other groups have performed hydrodynamical simulations of 
granulation over a grid of stellar parameters; \cite{magic13,magic14,trampedach13,tremblay13,beeck13,beeck13b}, but do not provide 
potentially useful values of the convective blueshift. 
\cite{magic14b} however found larger granules (usually associated with larger velocity fields) for hotter stars, consistent with a larger convective blueshift. In principle both effects (absolute and differential velocity shifts) can be extracted from numerical simulations providing they model enough spectral lines: this is discussed in Sect.~4. 

Concerning the second issue however, it is mostly uncharted and no previous study 
has estimated whether the 
response of the granulation to magnetic field depends on the spectral type or not. 
Several studies have aimed at 
measuring the temperature variations of individual stars during stellar cycles \cite[][]{gray92,gray94,gray95,gray96a,gray96}, yet 
no systematic tendency is available today.

In this paper we therefore focus on this last aspect. We aimed to precisely estimate the convective blueshift 
for a large sample of G-K main sequence stars but also measure how this blueshift varies with magnetic activity. 
To this end, we measured the differential velocity shifts of spectral lines, and more specifically their slope, defined in a robust way, 
in order to estimate the amplitude of granulation for each star.  After a description of our data analysis 
in Sect.~2, we derive the dependency on spectral type and activity level of the granulation amplitude using this criterion in Sect.~3. 
We focus on the temporally-averaged properties of stars: average activity level and convective level. 
Then in Sect.~4, we propose an estimation of the absolute shifts of spectral line bisectors from the criterion estimated in Sect.~3, using the Sun as a reference and adopting some assumptions. 
Finally, in Sect.~5 we reconstruct  $RV$ amplitudes as a function of 
the activity variability amplitude.  The $RV$ jitters as a function of 
{\it average} activity level have been studied in the past, but seldomly as a 
function of the amplitude activity amplitude. 
We conclude in Sect.~6.

\section{Data analysis}

\subsection{Data sample}

\begin{figure} 
\includegraphics{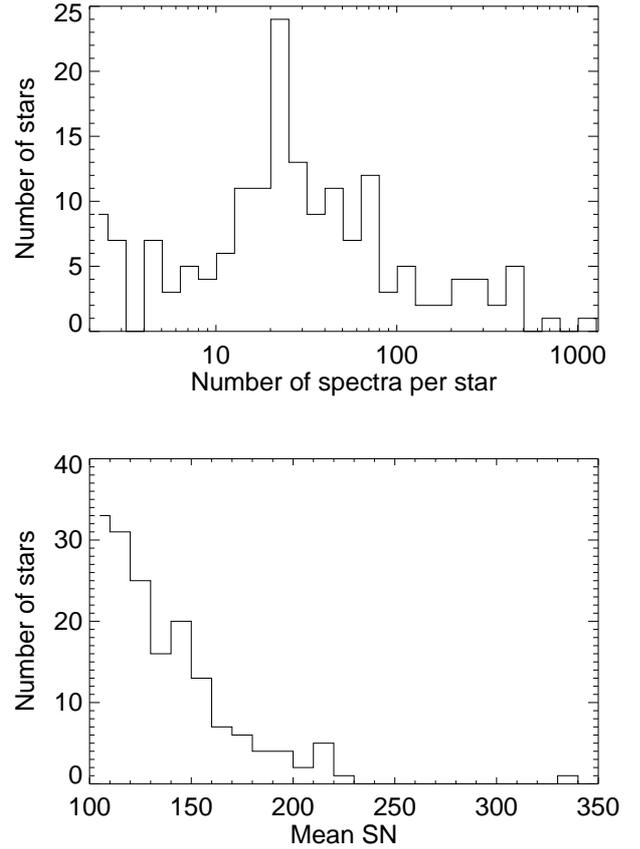}
\caption{
{\it Upper panel}: Distribution of the number of spectra per star. 
{\it Lower panel}: Distribution of the average S/N per star.
}
\label{sample}
\end{figure}

\begin{figure} 
\includegraphics{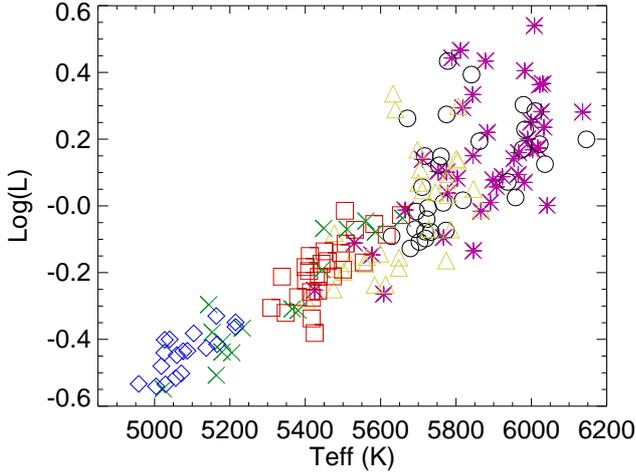}
\caption{Log of the luminosity (relatively to the solar luminosity) versus temperature (from Sousa et al 2008) for the  main sequence stars in our six samples: 
G0 (pink stars), G2 (black circles), G5 (yellow triangles), G8 (red squares), K0 (green crosses), and K2 (blue diamonds).
}
\label{hr}
\end{figure}

\begin{table}
\caption{Sample characteristics}
\label{tab_sample}
\begin{center}
\renewcommand{\footnoterule}{}  
\begin{tabular}{llllll}
\hline
Sp. T. & N$_{\rm Star}$ & N$_{\rm Spectra}$ & N$_{\rm Lines}$ & $<$Teff$>$ & $<$B-V$>$\\
\hline
G0 V     & 41 & 2664  & 152 & 5872 & 0.606 \\
G2 V     & 31 &  2386 & 148 & 5815 & 0.626 \\
G5 V     & 31 &  953  & 153 & 5665 & 0.678 \\
G8 V     & 27 &  2289 & 154 & 5456 & 0.734 \\
K0 V     & 16 &  660  & 148 & 5327 & 0.797 \\
K2 V     & 21 &  2136 & 169 & 5061 & 0.894 \\
\hline
Total  & 167 & 11088 & 196 & - & - \\ 
\hline
\end{tabular}
\end{center}
\tablefoot{Temperatures are from Sousa et al (2008). B-V are from the Simbad database at the CDS. }
\end{table}

We have extracted 167 stars, grouped into six samples from the HARPS survey described in \cite{sousa08}, and  covering 
the spectral types G0, G2, G5, G8, K0, and K2. The spectral types have been extracted from the Simbad database at the CDS\footnote{http://simbad.u-strasbg.fr/simbad/}. 
This survey was biased against very active and young stars, and focuses on 
stars with $v$sin$i$ lower than 3-4 km/s, as pointed out by \cite{lovis11b} in the study of cycles in stars from the same survey.

The spectra have been retrieved from the ESO archive\footnote{http://archive.eso.org/wdb/wdb/adp/phase3\_spectral/form}: We analyse 
1D spectra produced by the ESO Data Reduction Software after the interpolation of the 2D echelle spectra (one spectrum per order) over a grid with a constant step in wavelength (0.01~$\AA$). The original 2D spectra are also retrieved for the computation of the uncertainties. 

We only retained  spectra with a S/N, averaged over all orders,
larger than 100. Stars with less than three such spectra were eliminated. Indeed \cite{landstreet07} insists on the need 
to consider only high S/N spectra for this type of analysis, especially if a small number of lines is considered. 
Although the  S/N does not seem to bias 
our results, low S/N values lead to a large dispersion, which increases the uncertainties and  
decreases the number of spectral lines which can be analysed, hence the necessity for our threshold.
Fig.~\ref{sample} shows the distribution of the number of spectra per star and the average 
S/N per star. For most stars, we analysed between 10 and 100 spectra, the number ranging between 3 and more than 
1000. The S/N values typically lie between 100 and 200. 

Fig.~\ref{hr} shows the luminosity versus Teff for our six samples, provided by \cite{sousa08}. 
The global properties of each of these samples are shown in Table~\ref{tab_sample}. The B-V values have been retrieved from the CDS.
The individual values for each star are indicated in Table~\ref{tab_sample2}.

\subsection{Spectra correction and line identification}

\subsubsection{Continuum correction}

We first normalised the spectra for the continuum in two steps: we first identified an upper envelope for each spectrum and then applied a 
correction factor, to take into account the fact that due the presence of noise the actual continuum is slightly lower than this upper envelope. The procedure is described in Appendix B.  
We estimated that the resulting uncertainties on the intensities were usually lower than 1\%, and much better at high S/N. 

\subsubsection{Line identification}

We used a list of spectral lines of Ca, TiI, TiII, and FeII lines \cite[][]{dravins08}, and FeI lines   
\cite[][]{nave94}. Only some of these were selected for the final analysis (see 2.3). 
The lines were identified on each spectrum as local minima (over 41 pixels) below a flux equal to 90\% of the continuum. Note that for 
the purpose of line identification, we shift the spectra towards a zero shift position 
(with a one pixel of 0.01~$\AA$ precision) to ensure that the position of the minimum is as close as possible to the theoretical wavelength: in practice, this is done manually for one spectrum per spectral 
type which serves as a reference and all other spectra are shifted to match this reference spectrum. No 
interpolation is done, as the pixel precision is sufficient for the line identification and it does not affect the following analysis.


Before any computation on the spectral lines, we eliminated blended lines whose wavelengths were very  close to each other in our list 
and by doing a visual inspection.
Another selection using automatic criteria was then implemented (see next section).

\subsection{Differential velocity shifts of spectral lines}

To compute the differential velocity shifts of spectral lines, 
we performed a polynomial fit of the bottom of the line over five pixels 
(covering a range of 0.04~$\AA$, i.e. corresponding to  about 40\% of the typical line width at half maximum). We then estimated the position of the minimum of the fit and by comparison with the laboratory wavelength, computed the velocity of the line. 
For each line we therefore derived a wavelength and its uncertainty, which we then transformed into a radial velocity 
$RV$ (relative to the laboratory wavelength) and its uncertainty $\sigma_{\rm RV}$.
A similar method was also applied by \cite{allendeprieto02} and \cite{ramirez08}.
We also derived 
the flux at the bottom of the line, $F$, and its uncertainty $\sigma_{\rm F}$. 
We finally computed the bisector of each line, which was then used for the line 
selection. 
We note that in the red part of the optical spectrum, the presence of telluric lines may have impacted a priori our results: this is discussed in Sect.~2.5. 

We then selected the lines to be used for our analysis
using several automatic criteria on each line. 
We chose to eliminate: lines with a bisector slope outside the 3-$\sigma$ range of all bisectors 
\cite[this criteria is close to the one used by][but our threshold was not fixed]{ramirez08} or
a rms of the bisector fit residual outside the 3-$\sigma$ range; 
lines with a line width outside the 3-$\sigma$ range where $\sigma$ is deduced from the gaussian fit of the distribution of the variable.

In most of our computations (especially the slope of $RV$ vs $F$) we considered only 
strong enough lines, that is,  points with $F$ lower than 0.6 only, except 
when indicated otherwise. For selection purposes, we also add a criteria based on the 
$RV$ - $F$ relationship. For several lines, the $RV$ strongly deviates from the average 
$RV$ found using the other lines,
meaning that such lines could also be blended. We therefore eliminated the lines with measurements outside 
5-$\sigma$ (around the linear fit $RV$ versus $F$). 
Finally, lines appearing in only one star  of the stellar type sample were also eliminated. 
This allows us to apply a robust criterion on the  differential velocity shifts of spectral lines, as described in the 
following section. 
The final list of selected lines can be found in Table.~\ref{tab_line}.

\subsection{A criterion to characterise the differential velocity shifts of spectral lines}

\begin{figure} 
\includegraphics{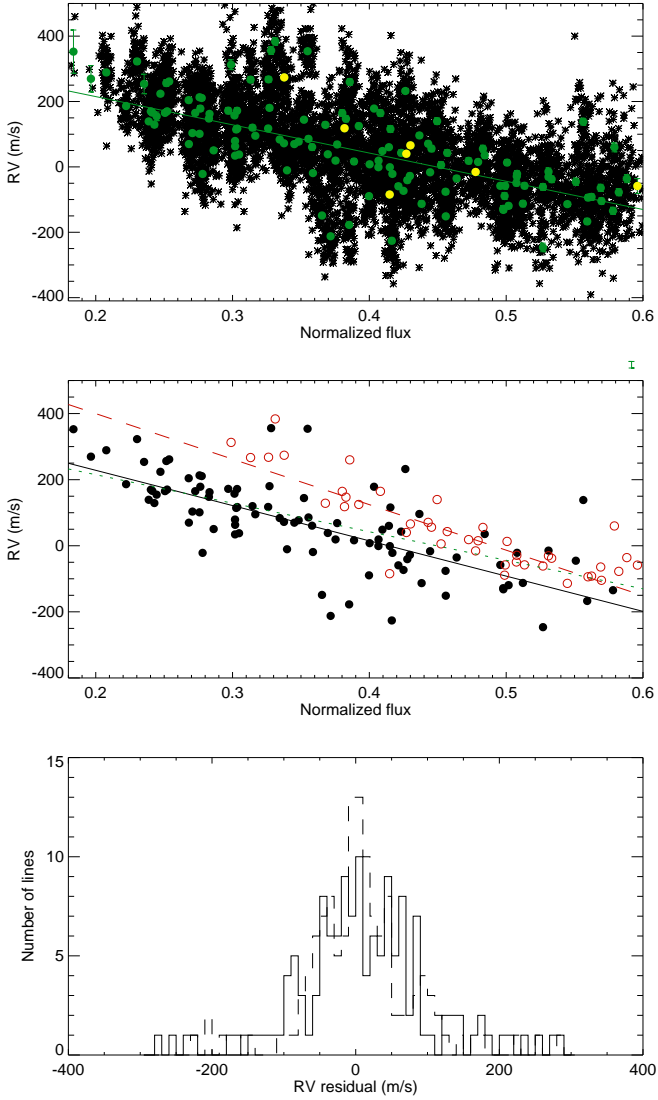}
\caption{
{\it Upper panel}: $RV$ versus a normalised flux of the bottom of the lines for HD223171 (G2), 
for lines deeper than the 0.6 threshold. Crosses represent individual measurements while green dots correspond 
to the temporal average for each line. 
The yellow dots are lines used by Gray (2009). The straight line is a linear fit on the green dots.  
{\it Middle panel}: Same as above but showing the wavelength dependence: The black filled circles and solid line are for lines with wavelength below 5750~$\AA$, 
while the red open circles and dashed lines are for wavelengths above 5750~$\AA$. The green dotted line is the linear fit on all points. 
{\it Lower panel}: Distribution of the $RV$ residual after removal of the linear fit, that is, the green dots of the upper panel (solid line). 
The rms $RV$ is 84 m/s. The dashed line shows the same distribution but for the residual computed after the wavelength-dependence correction 
(see text), and the rms $RV$ is 78 m/s.  
}
\label{ex_tss}
\end{figure}

An example of the differential velocity shifts of spectral lines obtained 
after the line selection described above is shown in Fig.~\ref{ex_tss} (upper panel) for 
the G2 star HD223171. 
Typical formal uncertainties on $RV$ values after temporally  averaging the $RV$ 
obtained using the same line are in the range 5-15 m/s, while the rms around the linear fit in that 
example is 84 m/s. 138 lines are used in this example.  
The average $RV$ computed at each time step over all selected 
lines has been substracted (the zero is therefore arbitrary here). 
A correction using the stellar $RV$ provided by the ESO Data Reduction Software does not change our subsequent results significantly.

The dispersion of the residual $RV$ is important. It can be due to several factors, for example: telluric lines, impact of a wavelength dependence, uncertainties on laboratory wavelength (all discussed in Sect.~2.5), impact of blends not taken into account or other properties of the lines, such as the 
excitation potential \cite[][]{dravins81}.

We defined a criterion to easily study the shape of the differential velocity shifts of spectral lines as a function of
various parameters:
we computed the slope of $RV$ versus $F$ values after averaging all points corresponding to a given
line (i.e. typically on the green points on Fig.~\ref{ex_tss}). We note that the formal uncertainties for $RV$ 
seem to be slightly under-estimated, as estimated by the comparison of the dispersion in $RV$ and $F$ for a given line and the formal uncertainties, by about a factor 1.3 on average but sometime by up to a factor 2: 
We therefore use the uncertainties 
derived from the dispersion (of the $RV$ obtained for a given line) to avoid underestimating our uncertainties.
This slope is named hereafter TSS, for `Third Signature Slope', taking up the term `third signature' proposed by \cite{gray09} to name the velocity shifts of spectral lines. 
An example of the linear fit corresponding to the TSS of HD223171 is shown 
in Fig.~\ref{ex_tss}).
In the following we focus on the analysis of the TSS computed over all spectra of a given star.

\subsection{Sources of dispersion or biases}

\subsubsection{Telluric lines}

Telluric lines caused by water may impact the line position estimation. Water wavelengths have been extracted from the HITRAN database 
\cite[][]{rothman13} available at http://hitran.org/. We listed the telluric lines which are within 0.02~$\AA$ of the lines
used in our analysis.
Most of these lines are very weak and are more than one order of magnitude lower than the stronger lines.
Out of the 197 final lines used to  compute the TSS in our selection, 
5.5\% have water telluric lines within 0.02~$\AA$ and the percentage decreases to 3\% 
when only considering lines that are used in 
all six samples (i.e. lines that are most likely to highly impact the TSS estimations). 
The residual of $RV$ versus $F$ after substracting the linear fit (used to determine the TSS) is not correlated with the 
position of telluric lines 
nor with their amplitude. We therefore estimate that the impact on our results should be negligable.

\subsubsection{Wavelength dependence}

Fig.~\ref{ex_tss} shows the $RV$ versus depth for two different wavelength ranges for HD223171. There is a clear shift between 
the two sets of points; lines at shorter wavelengths tending to show larger blueshifts. We observe this effect on most of our stars 
as well as for the Sun. Note that the difference 
between the average $RV$ of the black and red dots on the figure is of the order of 10 m/s only, while the shift in $RV$ between the two 
straight lines is one order of magnitude larger. Such an effect has already been identified by \cite{dravins81} and later by 
\cite{hamilton99}, using 298 FeI lines, and for the Sun in both cases. This effect is usually neglected in stellar studies \cite[][]{allendeprieto02} or not relevant because of the small 
wavelength coverage \cite[][]{gray09} but could explain some differences we observe with the differential velocity shift of \cite{gray09}, in particular the difference in shape which may be a wavelength effect and the much larger dispersion when considering more lines. 

We measured this effect for a large sample of stars for the first time. To quantity this effect, we proceeded as follows: 
we considered a simple linear relationship between wavelength and $RV$ and substracted from the $RV$ values a factor equal to  $p_\lambda \times \lambda$, 
where $p_\lambda$ is fitted to obtain a minimal rms of the $RV$ residual after the wavelength- and depth- dependance correction. The rms being extremely sensitive to outliers, we performed this computation 100 times for half of the measurements only (which were picked up randomly in the original data set), which also allowed us to derive uncertainties. We then considered the average of the 100 values to estimate $p_\lambda$ and derive the uncertainties from the distribution of the values.
We find values of $p_\lambda$ between -0.05 and 0.011~m/s/$\AA$ typically. 
The amplitude is compatible with the results of \cite{dravins81} and \cite{hamilton99} for the Sun. We conduct a detailed analysis of this effect in Sect.~3.3.

\subsubsection{Wavelength uncertainties}

As pointed out by \cite{dravins08}, laboratory wavelength uncertainties have a strong impact on $RV$ determination. The FeII and TiI laboratory wavelengths we used in this work may lead to imprecisions of the order of 6 m/s at, for example, 5000~\AA, which is small compared to the observed dispersion. However, most of the FeI wavelength uncertainties are probably one order of magnitude larger and may account for some of the observed dispersion \cite[][]{nave94,hamilton99}. For the example shown in Fig.~\ref{ex_tss}, the quadratic difference between the rms of the residuals for the two categories is 62~m/s.

\subsection{Activity level}

\begin{figure} 
\includegraphics{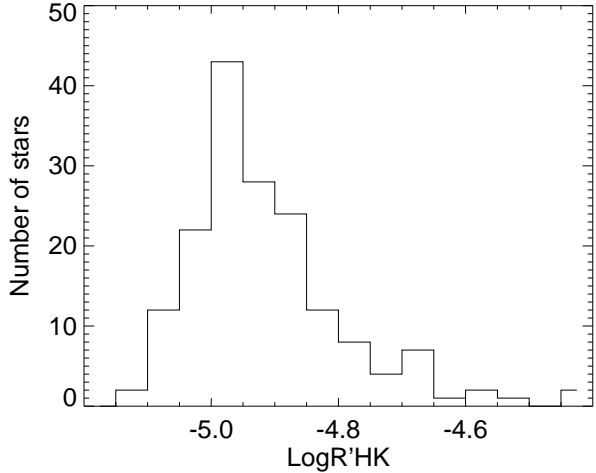}
\caption{ Distribution of the LogR'$_{\rm HK}$ values in our sample (one average per star).
}
\label{logrphk}
\end{figure}

The activity level for each epoch computed using the classical LogR'$_{\rm HK}$ derived from the flux in the Ca II H and K lines, 
gives the Log of the 
chromospheric emission. Our values have been corrected using the HARPS-Mt Wilson correcting factor  of \cite{lovis11b}. 
We did not introduce any correction for the thorium leaks, but we estimate that this should not impact 
our results significantly since we have selected spectra with a good signal to noise ratio and therefore 
with a large incoming flux. We have not taken into account the metallicity effect either. The distribution of 
the LogR'$_{\rm HK}$ values (averaged for each star) is shown in Fig.~\ref{logrphk}.

 \cite{lovis11b} studied 131 of the same stars included in our sample and there is a good correlation between the two estimates of LogR'$_{\rm HK}$. 
Our computation leads to values approximately 0.5\% smaller (i.e.a difference of the order of 0.025) than those derived by \cite{lovis11b}. This difference 
is not due to the difference in temporal coverage. 
Our LogR'$_{\rm HK}$ could be slightly biased due to the 
presence of thorium leaks or to the different spectrum selection, as well as to different B-V values. Nevertheless, this  does not impact our conclusion. 

\subsection{Analysis}

The analysis performed in Sect.~3 allows us to study the dependence of the TSS on 
both the spectral type (or B-V / Teff) and on the activity 
level. The TSS and averaged LogR'$_{\rm HK}$ values for each star are shown in Table~\ref{tab_sample2}. For each spectral type,
we compute the slope of the TSS versus LogR'$_{\rm HK}$. 


In a second part of the analysis (Sect.~4 and 5), we use the TSS to derive an average convective blueshift for each star.
We consider the Sun as a reference, using the solar spectra obtained by \cite{kurucz84} and 
reduced in 2005 by Kurucz\footnote{http://kurucz.harvard.edu/sun/fluxatlas2005/}. 
This allows us to derive $RV$ temporal variations for each star
following different assumptions, which may be compared to the actual $RV$ variations.



\section{Analysis of the differential velocity shifts of spectral lines}

\subsection{Dependence on B-V and Teff}

\begin{figure} 
\includegraphics{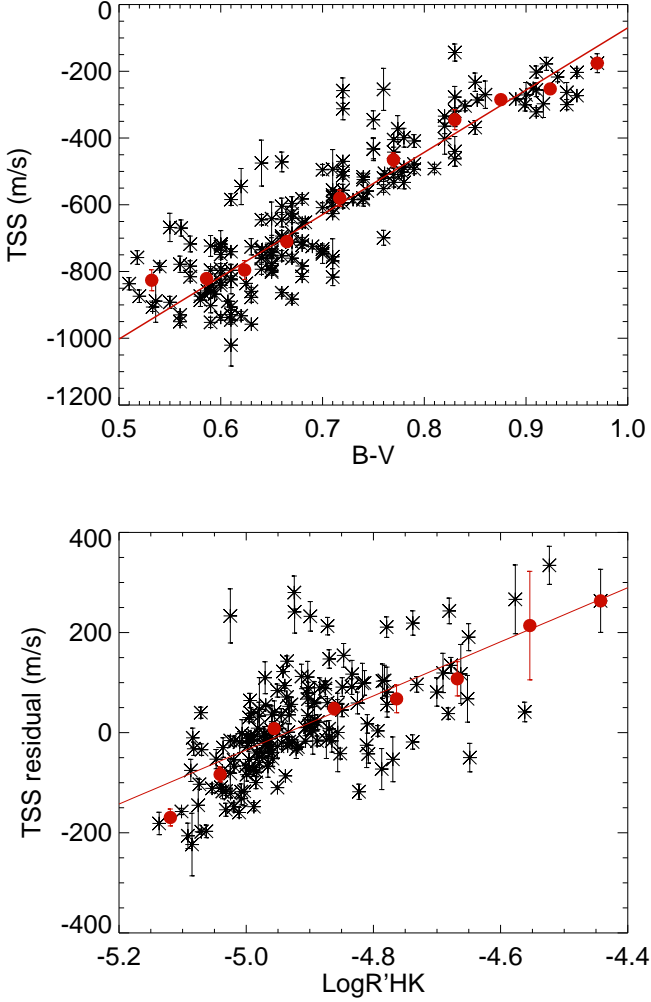}
\caption{
{\it Upper panel}: TSS (slope of the differential velocity shifts of spectral lines, in m/s/(F/Fc)) versus the B-V of the stars. The solid line indicates a linear fit while the red dots correspond to TSS averaged 
over bins in B-V. 
{\it Lower panel}: Residual of the TSS after correction of the B-V dependence versus LogR'$_{\rm HK}$. The solid line is a linear fit 
to the TSS. 
}
\label{bv_teff}
\end{figure}

The upper panel of Fig.~\ref{bv_teff} show the TSS versus B-V. The TSS charaterises the slope of the differential velocity shifts of spectral lines and
is strongly correlated with B-V (coefficient of 0.88, also meaning a strong anti-correlation with Teff).
If we extrapolate the TSS towards larger B-V (smaller Teff), it reaches zero for B-V=1.04 (and Teff=4680 K).
Fig.~\ref{bv_teff}  also shows the TSS averaged over bins in B-V: there may be a trend for a 
saturation of the TSS at low B-V, but this would need to be confirmed with a study of more massive stars.

\subsection{Dependence on activity}

\begin{figure} 
\includegraphics{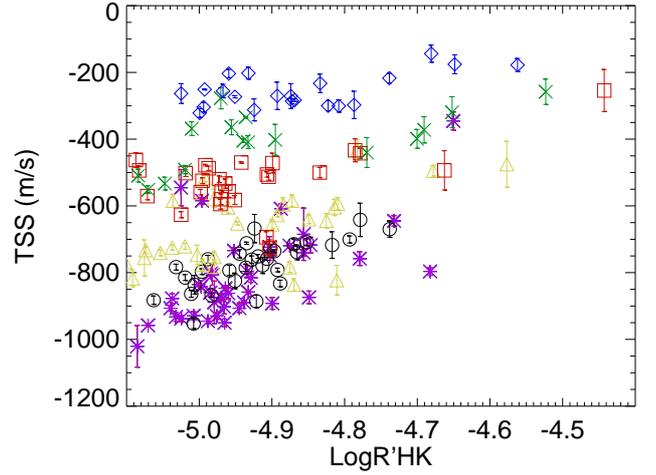}
\caption{
TSS (slope of the differential velocity shifts of spectral lines, in m/s/(F/Fc)) versus LogR'$_{\rm HK}$ for all stars in each sample (same colour and symbol codes as in Fig.~2). 
}
\label{global}
\end{figure}

Fig.~\ref{global} shows the TSS versus LogR'$_{\rm HK}$ for all stars, colour-coded depending on their spectral types.
We observe that for a given LogR'$_{\rm HK}$, |TSS| is above a minimal value which decreases as the activity level 
increases. This minimum value is around 1000~m/s/(F/Fc) (in the following the TSS is in m/s/(F/Fc) where F/Fc represents the flux normalised to the continuum) 
for the less active stars, and then decreases  towards a few hundred~m/s/(F/Fc)
for the most active stars in our sample.

After substraction of the B-V dependency derived in the previous section (using a linear fit), the TSS residuals are shown on the lower panel of Fig.~\ref{bv_teff}. 
These residuals are correlated with LogR'$_{\rm HK}$ (coefficient of 0.63), with a 
slope of 540$\pm$30 m/s/(F/Fc).
After correction of this LogR'$_{\rm HK}$ dependency, the rms on the TSS residuals is about 80~m/s/(F/Fc). 
It is still larger  than the formal errors on the individual TSS values (estimated when performing the linear fit
to compute the TSS and based on the $RV$ and $F$ 
uncertainties) by a factor of a few units, however, showing that other effects influence the TSS. 

For LogR'$_{\rm HK}$ lower than 
$\sim$-4.85, the slope of the residual TSS versus LogR'$_{\rm HK}$ is probably larger than the global slope, while for more active stars the trend may not
be as significant: there could be in fact two regimes, one in which the TSS is 
very dependent on the activity level (for the less active stars, with a slope of 872$\pm$39~m/s/(F/Fc)), and one in which the TSS is less dependent 
on activity (with a slope of 519$\pm$78~m/s/(F/Fc)), suggesting different properties of the magnetic fields. Note that in this description 
the Sun would be at the edge of the non-active 
star domain with LogR'$_{\rm HK}$ values between -4.95 and -4.85. 


We have noted in Sect.~2.6 that our LogR'$_{\rm HK}$ values were slightly smaller than those of 
\cite{lovis11b}, by approximately 0.5\%, with a large dispersion.
This may impact the slope TSS (or TSS residuals) versus LogR'$_{\rm HK}$. A difference of 
0.5\% impacts the slope with the same amplitude. If we recompute 
these slopes for each star sample, when there remains enough stars in common, 
we find larger slopes with differences between 0.5 and 20\% (although the number of stars 
is smaller): our conclusions are therefore not significantly affected by this possible discrepancy.

\subsection{Dependence on wavelength}

\begin{figure}
\includegraphics{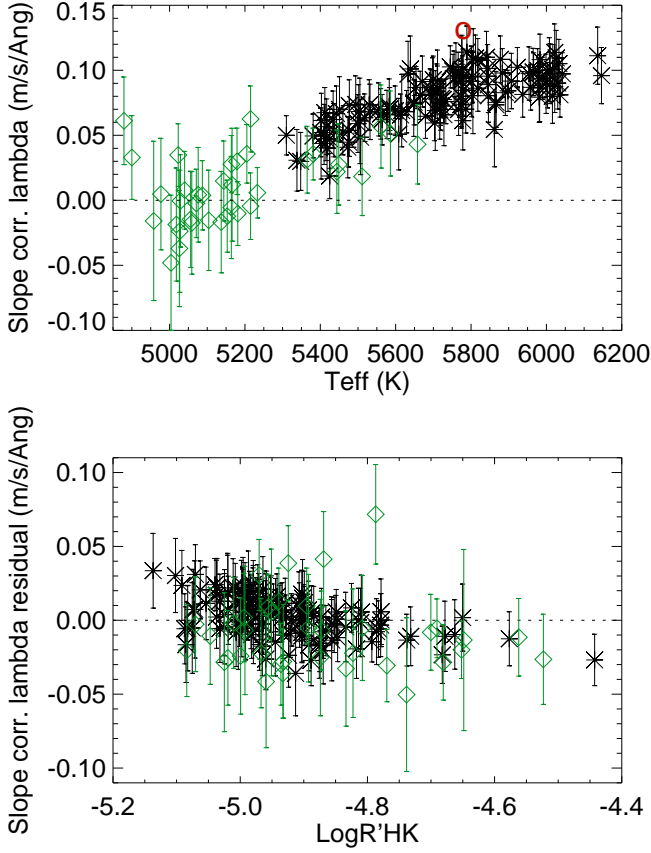}
\caption{
{\it Upper panel}: Slope of the wavelength dependence factor $p_\lambda$ of the differential velocity shifts versus Teff for G (black stars) and K (green diamonds) stars. The solar value of  $p_\lambda$ is 0.124 and is shown as a red circle. 
{\it Lower panel}: Same for the $p_\lambda$ residuals after correction of the Teff relationship versus LogR'$_{\rm HK}$. 
}
\label{pente_lambda}
\end{figure}

Fig.~\ref{pente_lambda} shows the $p_\lambda$ factor defined in Sect.~2.5.3 versus the TSS and Teff for our star sample. $p_\lambda$ characterises the amplitude of the dependency of differential velocity shifts on the wavelength.
We find a strong correlation between this factor and the amplitude of the TSS and with Teff,  at
least for Teff larger than 5400 K. 
The dispersion is, however, quite large for low Teff, which may be due to the very small signal in that domain.
$p_\lambda$ is equal to 0.124~m/s/$\AA$ for the solar spectrum: this solar value is compatible with the results in Fig.~\ref{pente_lambda} since it is very close to the values obtained for stars at the same Teff. 
The impact of this effect on the TSS is significant: for example it decreases from -863 m/s/(F/Fc) to -1163 
 m/s/(F/Fc) for HD223171 and from -799 m/s/(F/Fc) to -1107 m/s/(F/Fc) for the Sun. 

Finally, we substract the Teff dependence
from $p_\lambda$ and show the residuals versus LogR'$_{\rm HK}$ in Fig.~\ref{pente_lambda} (lower panel). The rms of the residuals is 0.017 m/s/$\AA$. We find that the activity 
level impacts the maximum value which can be taken by the slope so that high activity levels reduce the wavelength dependence, at  least for G stars. 

Both the Teff and the LogR'$_{\rm HK}$ impacts on $p_\lambda$ represent strong constraints for the simulation of stellar convection over a large grid of stars, with or without magnetic fields. The wavelength dependence is likely to be due to the stronger contrast of granulation towards lower wavelengths, reinforcing the convective blueshift. This interpretation agrees with our observed trends as a function of Teff and activity, and such measurements may provide constraints on granulation contrasts. 


\subsection{Analysis for each spectral type}

\begin{table}
\caption{Main results: TSS and activity and wavelength dependencies}
\label{tab_res}
\begin{center}
\renewcommand{\footnoterule}{}  
\begin{tabular}{llllllll}
\hline
Sample & TSS & $\sigma$TSS & Slope & $\sigma$Slope  & $p_\lambda$   \\
       &     &             &       &                &               \\
       & (m/s & (m/s     & (m/s & (m/s          & (m/s/$\AA$)         \\
       & /(F/Fc))& /(F/Fc))      & /(F/Fc)) & /(F/Fc))          & (m/s/$\AA$)         \\
\hline
G0 &-811.7& 20.8  &829.5 &44.0 &  0.092 \\
G2 &-774.1& 12.9  &672.3 &57.7  & 0.086 \\
G5 &-672.3& 19.7  &466.8 &56.6 &  0.077 \\
G8 &-523.2& 17.2  &349.0 &62.3 &  0.056 \\
K0 &-402.1& 22.6  &318.4 &49.8  & 0.026 \\
K2 & -254.3& 11.2 &231.0 &40.0 &  -0.006 \\
\hline 
\end{tabular}
\end{center}
\tablefoot{For each spectral type there is shown: the TSS (without the wavelength correction), the slope of TSS versus LogR'$_{\rm HK}$ associated to its uncertainties and the wavelength factor $p_\lambda$. The TSS is the slope of the differential velocity shifts of spectral lines.
}
\end{table}

\begin{figure} 
\includegraphics{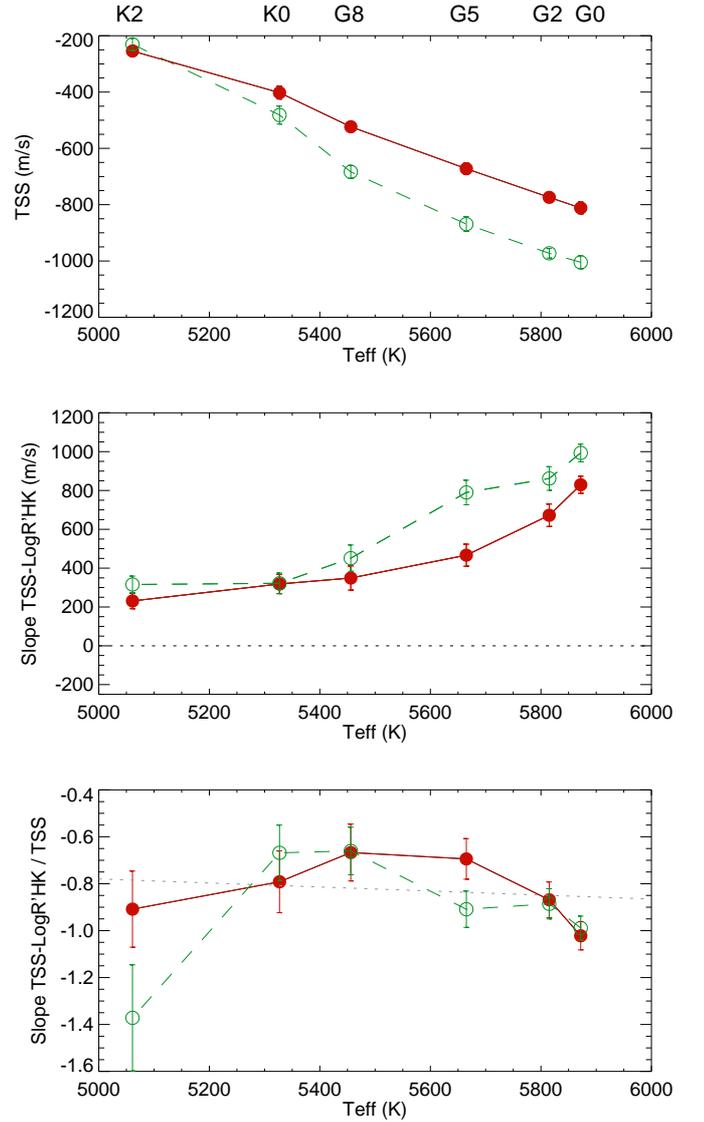}
\caption{
{\it Upper panel}: TSS (slope of the differential velocity shifts of spectral lines, in m/s/(F/Fc)) averaged for each sample versus the average temperature of the sample (red filled circles and solid line). 
The green curve (open circles and dashed line) 
shows the TSS after the wavelength dependence correction. 
{\it Middle panel}:  Same for the slope of the TSS versus LogR'$_{\rm HK}$.
{\it Lower panel}:  Same for the ratio between this slope and the TSS. The dotted line is a linear fit on the ratio in red.
}
\label{tss_type}
\end{figure}

The average TSS was computed for each of the six star samples. The results are shown in Table~\ref{tab_res}, 
and on Fig.~\ref{tss_type} (upper panel). 
There is a clear decrease of the TSS towards cooler stars, as seen above.
The TSS corrected for the wavelength dependence is also shown and appears larger in amplitude but shows a similar trend.

\begin{figure} 
\includegraphics{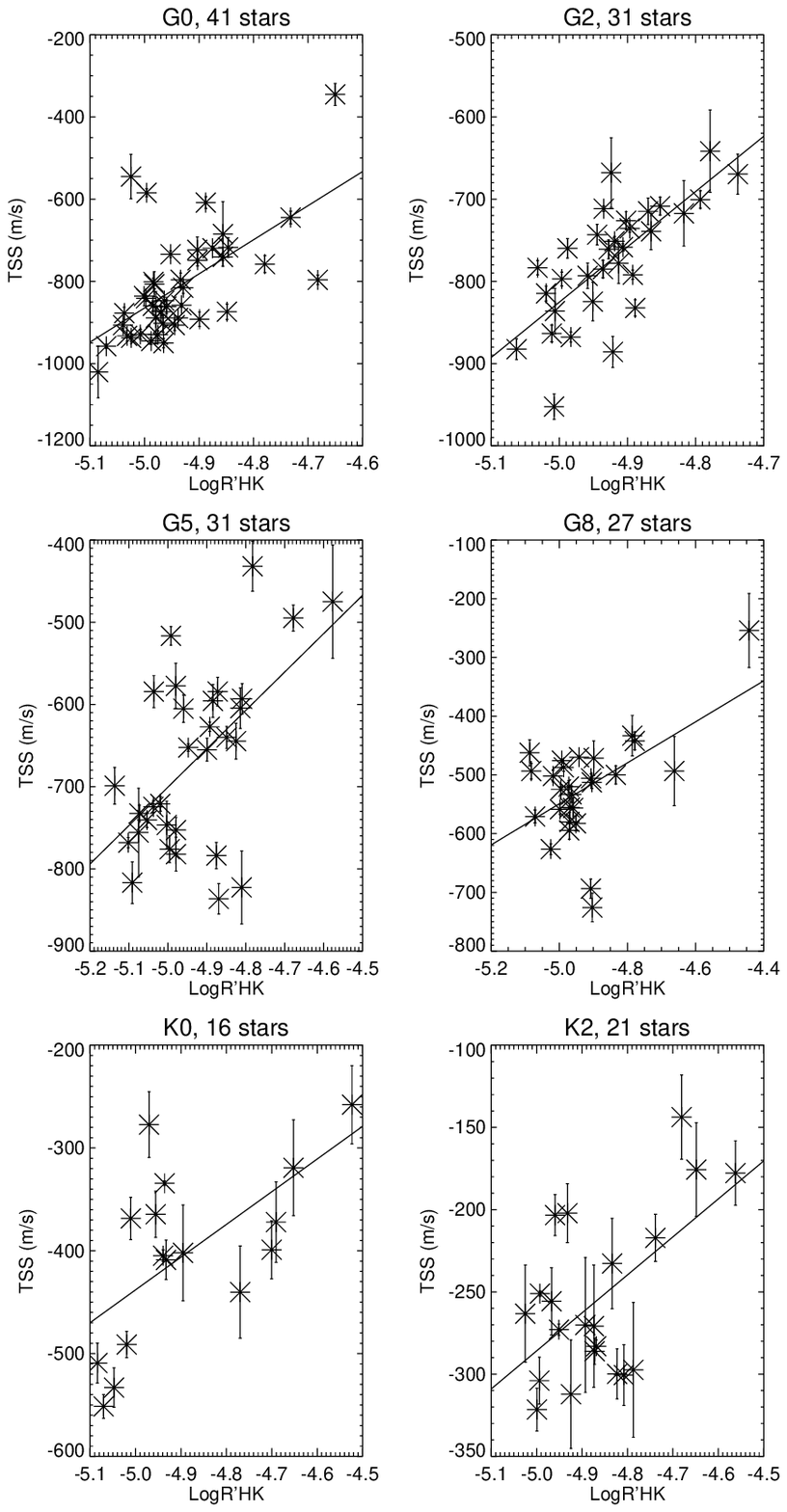}
\caption{Each panel shows the TSS (slope of the differential velocity shifts of spectral lines, in m/s/(F/Fc)) computed for each star versus the average LogR'$_{\rm HK}$ of that star. The straight line is a linear fit on the measurements.   
}
\label{res_type}
\end{figure}

We now consider the dependence of the TSS on the activity level separately for each spectral type in our sample. 
The TSS versus the average LogR'$_{\rm HK}$ is shown in Fig.~\ref{res_type},
with a positive correlation in all cases.
There are  however some deviations from the linear fit.  This could be due to the fact that the temperatures of the stars vary within a given sample, as well as to inclination effects (see Sect.~4.3 for a discussion). The possible uncertainty in spectral classification could also lead to some dispersion or outliers if some stars are in fact subgiants of luminosity class IV, or have been incorrectly classified. Stellar rotation may also impact our estimates \cite[as discussed for example by][]{ramirez09}: after taking into account the fact that both the TSS and the rotation rates vary with spectral type, we find that the TSS residuals do not show any significant trend versus $v$sin$i$ for our sample limited to $v$sin$i$ lower than 5 km/s. We do not, therefore, expect a strong bias in this $v$sin$i$  range. On the other hand, we also expect the differential rotation to impact the TSS, as shown by \cite{beeck13} for example, and believe that this effect, though difficult to quantify, could add some dispersion.

The slope of the TSS versus the average LogR'$_{\rm HK}$ is shown  
in Fig.~\ref{tss_type} (values are indicated in Table~\ref{tab_res}) and is smaller for cooler stars while it can be as large as 800 m/s/(F/Fc) 
for G0 stars. 
This corresponds to an amplitude of 80 m/s/(F/Fc) for an amplitude of 0.1 of the LogR'$_{\rm HK}$ 
(i.e. a typical difference in LogR'$_{\rm HK}$ between solar cycle minimum and maximum). 
An extrapolation of the slope with Teff gives a zero TSS for Teff=4830 K, which is close to the values derived for the TSS.  

Finally, the ratio between this slope and the TSS (i.e. between the curves from panels 1 and 2 of Fig.~\ref{tss_type}) 
is shown on the lower panel: this ratio, showing the attenuation factor of the convective blueshift due to activity, does not show any strong trend, suggesting that stars of 
different types probably have a very similar response to magnetic activity relative to the
amplitude of convection, that is, the convection may be reduced within the same proportions.

 We can compare this last result with the simulations made by 
\cite{beeck15}. Their hydrodynamic simulation of granulation in the presence of small scale magnetic fields for stars from F3 to M2 
shows a magnetic field strength independant from the spectral type, so that if more flux is available, it is spread over a larger surface. Such a behaviour has also been observed by \cite{steiner14} on a small range of simulation parameters.
In such conditions we would expect the response in terms of granulation properties (and hence the convective blueshift)  
to be similar from one spectral type to another, since the magnetic field properties do not change much, which is what we observe.

\section{Convective blueshifts}

In Sect. 1 we mentioned two related effects of convection on spectral line shifts: the absolute convective shift, which is relevant for the exoplanet $RV$ analysis for example ($RV$ typically computed over all available spectral lines), but difficult to measure directly, and the differential shifts measured in the previous section. To derive the former from differential shift measurements such as the TSS, we need to make some assumptions and use the Sun as a reference.

\subsection{Computed convective blueshifts}

\begin{figure}
\includegraphics{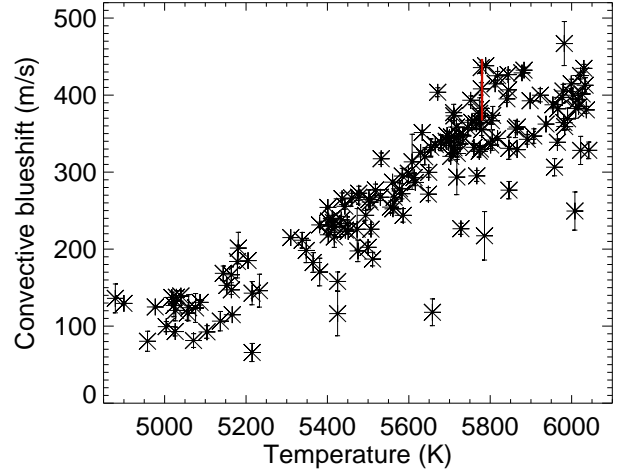}
\caption{
Reconstructed convective blueshift versus Teff for all stars in our sample.
The vertical red line corresponds to the solar convective blueshift derived from Reiners et al. (2016) for no weighting and for a weighting equal to the line depth.
}
\label{conv_bl0}
\end{figure}

\begin{figure} 
\includegraphics{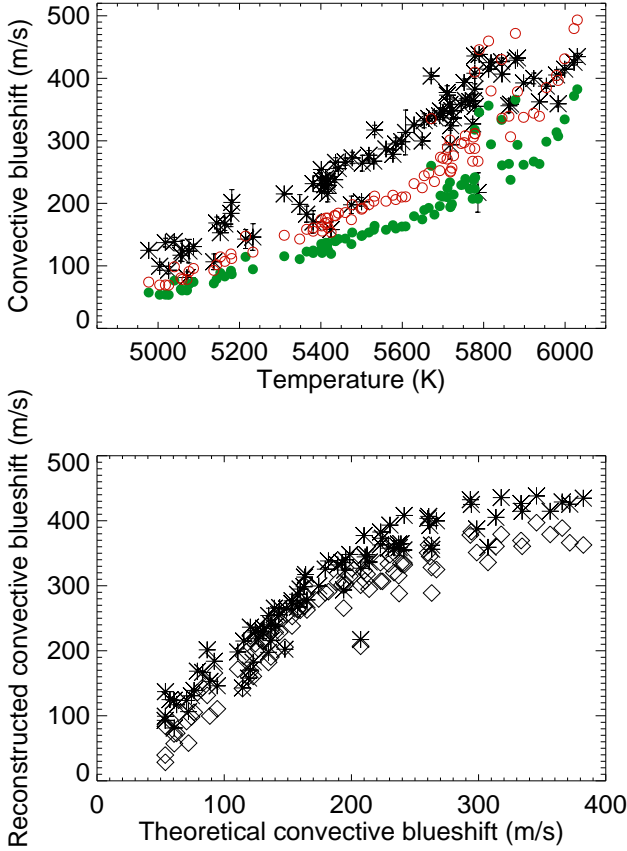}
\caption{
{\it Upper panel}: Reconstructed convective blueshift versus Teff for 
the stars for which the formula of Allende Prieto et al. (2013) could be 
applied (stars). 
The convective blueshift derived from Allende Prieto et al. (2013) for 
those stars is shown before correction from the wavelength range (red circle) and after correction (green dots).
{\it Lower panel}:  Reconstructed convective blueshift versus those derived from Allende Prieto et al. (2013), from the original TSS (stars) and those corrected  for the wavelength dependence (diamonds).
}
\label{conv_bl}
\end{figure}
                                                                                                                                                 
In this section, we use the TSS computed for each star to estimate its convective blueshift, using two principles:
\begin{itemize}
\item{We use the assumption made by \cite{gray09} that the shape of the differential shift of spectral lines is representative of the absolute convective blueshift due to granulation that we are interested in. In this same publication this factor is computed between the curve for a given star and the solar curve used as a reference, and is also used to derive the intrinsic radial velocity  of these stars: their results show a proportionality between the shape, or in our case the slope, and the absolute convective blueshift. In principle it would be possible to check such an assumption with output from numerical simulations such as those provided by \cite{ramirez09} for K stars or by \cite{magic14} for a larger range of stellar parameters, since they compute shifts from a large set of spectral lines leading to both differential and absolute values. Unfortunately, however, they do not attempt to check that assumption.
}
\item{We use the Sun as a reference. For that purpose, we need two solar values: the solar TSS$_\odot$ and the convective blueshift $RV_{\rm convbl \odot}$. We derive the solar TSS$_\odot$ from the solar spectrum of \cite{kurucz84} using the set of lines we used for the G2 sample. However, the spectral resolution of this spectrum is 500,000, that is, approximately four times larger than the resolution of HARPS spectra (approx. 120,000): \cite{hamilton99} pointed out that a degraded resolution has a significant impact on slopes such as the TSS we derive for resolution below 200,000, leading to smaller slopes (approx. 25\% for the HARPS resolution). We have therefore also computed TSS$_\odot$ on a degraded solar spectrum, which leads to TSS$_\odot$=-776~m/s/(F/Fc) \cite[the difference with the original value being smaller than what was derived by][]{hamilton99}. We use the value of -776~m/s/(F/Fc) in the following. 
As for the solar convective blueshift, a value of 300~m/s \cite[][]{dravins99} is often used \cite[for example in][]{meunier10}. 
It corresponds to the values measured over all lines taken into account in classical $RV$ computations such those commonly performed. This estimation is however uncertain, for example due to uncertainties in laboratory wavelengths. \cite{reiners16} have recently reevaluated an absolute $RV$ versus spectral line depth for the Sun (i.e. similar to what we have done in Sect.~3 but for absolute- rather than relative $RV$ ). Unfortunately, they do not provide the average convective blueshift that corresponds to their results. Using their relationship between the absolute $RV$ and line depth, and a  sample of spectral lines we extract from the solar spectrum of \cite{kurucz84} for depths between 0.05 and 0.95 and wavelengths between 4000 and 6600 \AA, we obtain an average convective blueshift of 435 m/s (i.e. larger than the previous value of 300~m/s). If we assume that when computing the $RV$ using cross-correlations between spectra, the $RV$ is more sensitive to deep lines (we assume a weighting factor equal to the line depth), we obtain a convective blueshift of 355 m/s. In the following we consider this the value to compute the convective blueshift. It should be multiplied by 1.22 if we wish to consider a convective blueshift based on average-between-spectral-line positions rather than a weighted one.
}
\end{itemize} 

 The convective blueshift of each star is therefore computed as follows:
\begin{equation}
    RV_{\rm convbl} = {\rm TSS} \times  RV_{\rm convbl \odot} / {\rm TSS}_\odot
,\end{equation}
where ${\rm TSS}$ is the stellar value derived in Sect.~3 and the solar values are as discussed above.
Fig.~\ref{conv_bl0} shows the resulting convective blueshifts for all stars in our sample versus Teff. There is naturally a strong correlation since our convective blueshift is proportionnal to the TSS. There may also be a plateau for Teff larger than 5800 K. Note that the convective blueshifts derived from the TSS corrected for the wavelength effect (not represented on the figure) are only 38~m/s lower than the original ones, so the effect is small.  

\subsection{Comparison of the adopted convective blueshift with theoretical results}

We now wish to compare our computed values with theoretical results. 

\cite{allendeprieto13} performed hydrodynamical simulations of granulation for various star parameters to deduce a number of properties for the exploitation  of GAIA observations, including a numerical expression of the convective blueshift of stars depending on their Teff,  log g and metallicity. The formula only being valid for certain ranges in parameters, we have only been able to apply their formula to 93 stars in our sample.  The red dots in Fig.~\ref{conv_bl} represent the theoretical 
convective blueshift for these stars based on the stellar parameters of \cite{sousa08}, while our estimation is represented by the black stars. This convective blueshift corresponds to the GAIA wavelength range, that is, 8470-8740~\AA. If we compute the solar convective blueshift from the results of \cite{reiners16} as above but selecting only lines in this wavelength range, we obtain 459~m/s (computation with line depth weighting) instead of the 355~m/s obtained above. Therefore, to be representative of a large wavelength range $RV$ computation, this theoretical convective blueshift should be divided by 1.29, as shown by the green dots. On the other hand, applying the formula of \cite{allendeprieto13} to the solar parameters leads to a convective blueshift of 285~m/s, which is significantly smaller than 459~m/s deduced from the observation of \cite{reiners16} for the same wavelength range suggesting that the numerical simulations produce a convection significantly weaker (by a factor of approximately 1.7 below 5800~K) than that observed. 

When comparing the simulated convective blueshift (green dots) with the estimated convective blueshift  based on the assumption of Sect.~4.1, we naturally find a significant difference in amplitude as well, since the Sun was used as a reference. That said, the trend for Teff lower than 5800~K is similar, except for the multiplying factor (a factor of approximately 2). Above 5800 K, the trend seems different, with a strong increase observed in the simulation but not in our observations: this will need to be investigated in the future.


\cite{magic14} also provide line shifts versus Teff from numerical simulations of stellar convection as well, with a similar trend in Teff. Their line shifts are 
approximately 100 m/s at 5000~K (similar to our values) but in the range 700-800 m/s at 6000~K, that is, slightly larger than our values and much larger than those 
of \cite{allendeprieto13}. This could be due to the fact that their line shifts are computed from the bottom of the line (and not for 
the whole line, as is assumed in the convective blueshift definition here), and/or may correspond to the specific fictitious Fe I line they are using meaning that a direct comparison of the amplitudes is difficult. 
An interesting feature observed by \cite{magic14} however is a saturation appearing at temperatures larger than 6000 K, which may be similar to what 
we observe.

\section{RV variability}

\begin{figure} 
\includegraphics{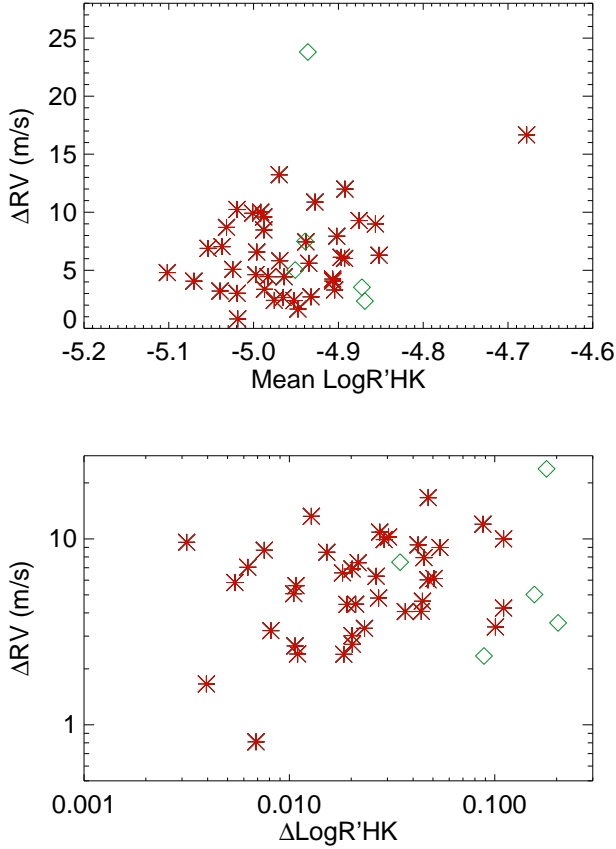}
\caption{
{\it Upper panel}: $\Delta RV_{\rm obs}$ versus the average LogR'$_{\rm HK}$, 
computed on the binned time series and for the 43 stars with more than two bins (see text), for G (red stars) and K (green diamonds) stars.
{\it Lower panel}: Same versus $\Delta$LogR'$_{\rm HK}$
}
\label{rv}
\end{figure}

We now wish to compare the observed $RV$ variability (derived in Sect.~5.1) with those which can be predicted using our estimated TSS (in Sect.~5.2).
We have thus retrieved the radial velocities computed by the ESO Data Reduction Software 
from the headers of the archive files. These $RV$ must be corrected for two effects: 
\begin{itemize}
\item{36 stars in our sample bear planets according to the 
Extrasolar Planets Encyclopaedia\footnote{http://exoplanet.eu/}. We therefore retrieved the exoplanet parameters, 
fitted the parameters when not published (this is sometimes the case for the time at periastron and
periastron argument), reconstructed the corresponding $RV$ time series, and finally substracted the planetary signal 
from the 
measured $RV$. Note that some of these stars bear several planets (22 with 1 planet, 9 with 2 planets, 3 with 3 planets and 1 with 4 planets). 
One of them has not been corrected because the three planets are all below 1 m/s. 
This is a necessary step, otherwise the induced $RV$ is 
overestimated and in some cases possible correlation with the LogR'$_{\rm HK}$ may be masked by the presence of the 
planetary signal.  In most cases the parameters have been retrieved from the compilation made by \cite{mayor11}. Others were taken from elsewhere \cite[][]{butler06,naef07,pepe11,hinkel15,diaz16}, and parameters not provided have been fitted on our time series.}
\item{A very strong $RV$ trend with time is exhibited by six particular stars suggesting the presence of a binary. We have removed the trend before analysing their 
$RV$ (using a linear fit or a second degree polynomial fit).}
\end{itemize}
 
\subsection{Observed $RV$ versus LogR'$_{\rm HK}$}

When only considering stars with at least 10 observations (131 stars), 40\% of the stars have a correlation between $RV$ LogR'$_{\rm HK}$ above 0.4 and 
24\% above 0.6. 
To compute long-term amplitudes, we averaged the $RV$ values in 50 day bins. We made sure to have at least 5 points 
in each bin (otherwise the observation was discarded) and considered stars with at least 4 such bins in the following analyses. This reduced our sample to 
43 stars.
From these time series we computed the $RV$ amplitude (defined as the maximum minus the minimum of the binned time series), 
$\Delta RV_{\rm obs}$,  as well as $\Delta$LogR'$_{\rm HK}$ and the average LogR'$_{\rm HK}$ for comparison purposes. 
$\Delta RV_{\rm obs}$ is compared in the next section with $RV$ variations derived from the convective blueshift. 
Considering stars with more bins considerably reduces the sample, although the statistics, in terms of stars with a good correlation between 
$RV$ and LogR'$_{\rm HK}$ , are relatively robust: 40\% of the stars have a correlation between $RV$ and LogR'$_{\rm HK}$ (from the binned series). 
The uncertainties on each $RV$ measurement as computed by the ESO Data reduction Software are between 0.2 and 0.6 m/s depending on the star. Because there is also some stellar  intrinsic variability and since some bins contain as few as five points, the uncertainties on $\Delta RV$ are larger, typically of the order of 1-2~m/s.

Fig.~\ref{rv} shows $\Delta RV_{\rm obs}$ versus the averaged LogR'$_{\rm HK}$  for each of the 43 stars (which shows no particular trend) and versus 
the amplitude of the LogR'$_{\rm HK}$. 
We see that strong variations in $RV$ tend to be associated with stronger amplitudes in 
LogR'$_{\rm HK}$, although there is a large dispersion. 

Our results can be compared with previous $RV$ variations published in the literature. Most of the time only the rms of $RV$ (before binning)
versus the average of the LogR'$_{\rm HK}$ or R'$_{\rm HK}$ (and not the amplitude, which we believe is more relevant) is available. 
Our rms $RV$ falls well within the range of variation of \cite{wright05} and \cite{isaacson10}, although we also observe stars with smaller 
rms. 
Furthermore, we find, as they do, that there is no clear trend when considering the rms versus the average R'$_{\rm HK}$. 
This is different from the results of \cite{santos00} and \cite{saar98}, who found larger rms and a trend. 
We have also computed the slope of R'$_{\rm HK}$ versus $RV$ (before binning) as studied by \cite{lovis11b} as a function of Teff. 
Our slopes are of similar amplitudes in the different Teff domains. 
 
The Sun, with an amplitude of logR'$_{\rm HK}$ of 0.1 and a predicted amplitude in $RV$ of 8 m/s \cite[][]{meunier10a} falls within the 
range of values found for our sample.

\subsection{Simulated $RV$}

\subsubsection{Computation of the simulated $RV$}

\begin{figure} 
\includegraphics{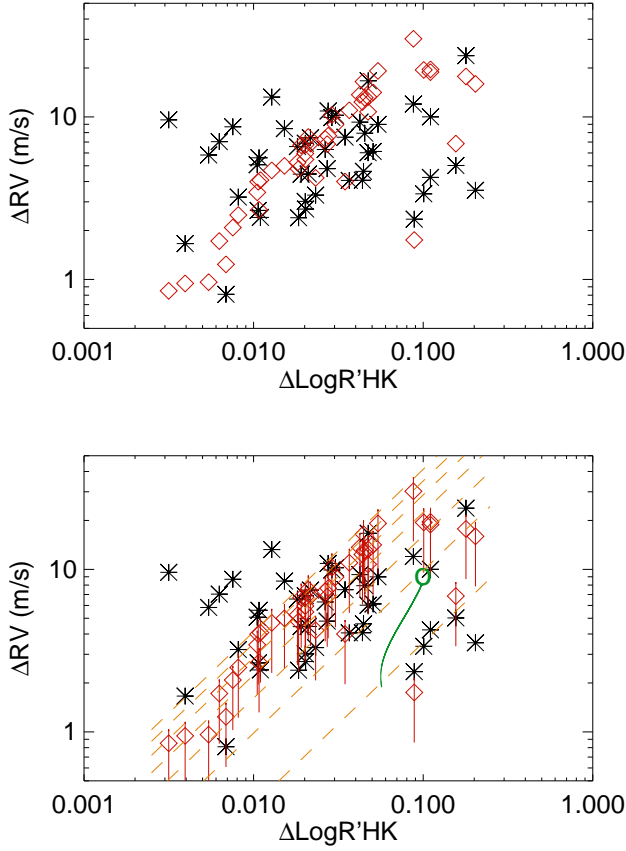}
\caption{{\it Upper panel}: $\Delta RV_{\rm obs}$ versus the amplitude of the LogR'$_{\rm HK}$, on averaged time series (see text), in black. 
$\Delta RV_{\rm conv}$ is shown in red.
{\it Lower panel}: Same as the upper panel. Red vertical lines correspond to the $\Delta RV_{\rm conv}$ reconstructed for inclinations of the rotational axis  
in the range 10--90$^{\circ}$. The green circle is the solar value (seen equator-on), and the green line indicates what  the solar 
value would be for various inclinations. The orange dashed line corresponds to $\Delta RV_{\rm conv}$ reconstructed from eq. 5 for Teff between 
4950 K (lower curve) and 6150 K (higher curve) with a step of 200 K between each curve. 
}
\label{rv_conv}
\end{figure}

We now estimate the $RV$ variations that would be due to the attenuation of the convective blueshift from the typical dependence of the convective 
blueshift on activity and from the amplitude of variation of the logR'$_{\rm HK}$ for each star.  
For a given spectral type, we defined in Sect.~3.4 the slope of the TSS versus LogR'$_{\rm HK}$, hereafter referred to as $G$. 
Hence, if a certain star activity level varies with $\Delta$LogR'$_{\rm HK\circ}$ then we expect a variation of 
$\Delta$TSS$_{\rm \circ}$ 
of the TSS, the ratio being the slope :

\begin{equation}
G = \Delta {\rm TSS}_{\rm \circ} / \Delta LogR'_{\rm HK\circ}
.\end{equation}

For this star, following eq. 1, the convective blueshift is:

\begin{equation}
 RV_{\rm convbl \circ} = {\rm TSS}_{\rm \circ} \times RV_{\rm convbl \odot} / {\rm TSS_{\rm \odot}} 
.\end{equation}

As a consequence, for any star we study with the same temperature (i.e. corresponding to the same $G$) and with an observed variability of 
$\Delta$LogR'$_{\rm HK*}$, we expect the TSS to vary by:

\begin{equation}
\Delta {\rm TSS}_{\rm *} = G \times \Delta LogR'_{\rm HK*} 
.\end{equation}

In the following, $R$ will be interpolated for each Teff from the curve in Fig.~\ref{tss_type}.
From eqs. 1 and 4, the variation of the convective blueshift with time (i.e. the $RV$ variation due to the attenuation of the 
convective blueshift) is:

\begin{equation}
\Delta RV_{\rm conv} = G \times  RV_{\rm convbl \odot} / {\rm TSS_{\rm \odot}} \Delta LogR'_{\rm HK}
.\end{equation}

$\Delta RV_{\rm conv}$ is then compared with $\Delta RV_{\rm obs}$ (derived in Sect.~5.1) 
and plotted as a function of $\Delta$LogR'$_{\rm HK}$  
in Fig.~\ref{rv_conv} (upper panel). The global amplitude corresponds to observations. We note that 
$\Delta RV_{\rm conv}$ is not correlated with $\Delta RV_{\rm obs}$ however, although the amplitudes are compatible, 
meaning that for a given star the reconstructed value can be quite different from the observed one. 
This could be due to the fact that eq. 5 does not include any dependence on inclination of the stellar rotational axis and therefore we assume in the following that 
it corresponds to an inclination of 45$^{\circ}$, as the slope $G$ for a given spectral type has been computed for a sample of stars with 
various random inclinations.

\subsubsection{Impact of inclination on the estimation}

The way we estimated the impact of inclination is detailed in Appendix C. 
The results are shown in Fig.~\ref{rv_conv} (bottom panel) as red lines. 
When taking into account the impact of inclination on our reconstructed $\Delta RV_{\rm conv}$, the obtained dispersion and range of values
correspond well to the observations. Note that the ranges have been computed assuming solar activity patterns: for a star with 
a different latitude distribution of magnetic activity (i.e. close to the equator, or on the contrary able to extend more poleward) we expect the 
laws such as those shown in Fig.~\ref{illust_inc} to be slightly different, leading to slightly different ranges. 
We conclude that a large part of the observed dispersion can be explained by the various inclinations of the stars in our sample.

For comparison purposes, Fig.~\ref{rv_conv} also shows  the curves corresponding to eq. 5 for seven values of
Teff spanning our sample (orange dashed line), starting with 4950 K (bottom line) and then with increasing Teff (with a step of 200~K) up to 6150 K. 
The green circle indicates the position of the Sun, and its track for various 
inclinations as computed by \cite{borgniet15}. The Sun is localised at a relatively low position compared to other stars of similar temperature and to the straight line corresponding to 5750~K. However, the stars in our sample which fall within the same Teff range all exhibit $\Delta$LogR'$_{\rm HK}$ lower than 0.05, therefore it is not straightforward to extrapolate their behaviour for a larger variability such as the one observed for the Sun. We conclude that inclination allows for a large dispersion in $RV$ amplitude for a given activity variability.

\section{Conclusion}

We have defined a criterion to characterise the differential velocity shift of spectral lines for a sample of 167 main sequence stars with spectral types in 
the range K2 to G0. We estimate the slope of $RV$ versus the flux $F$ at 
the bottom of each spectral line \cite[studied by][for the Sun]{dravins81}, which we named the TSS, 
versus the spectral type, the activity level and wavelength. We focus on temporally averaged properties. Our conclusions are as follows:

\begin{itemize}
\item{We find a decreasing TSS and therefore a decreasing convective blueshift amplitude with decreasing temperature, as expected, with a convective blueshift of approximately  150~m/s for K2 stars and 500~m/s for G0 stars. 
The derived convective blueshift based on the assumption made by \cite{gray09} that the convective blueshift is proportionnal to the shape 
(the slope in our case) of the differential velocity shifts shows a trend with temperature which is a in close agreement with the granulation simulations of \cite{allendeprieto13} but twice larger in amplitude for temperatures lower than 5800~K.  
There is also a discrepancy in the trend for the  highest temperature (above 5800K), suggesting further analysis is needed, as there may be a saturation effect 
not visible in the simulation by  \cite{allendeprieto13} but  obtained by \cite{magic14}.
}
\item{We find, for the first time, a significant and strong variation of the TSS (and hence of the convective blueshift) with the average activity level
of the star in the K2 to G0 domain. 
}
\item{The relative variations of the TSS with activity, that is, the slope of the TSS variation with LogR'$_{\rm HK}$ divided by the average TSS for that spectral type,  are relatively constant with spectral type. Therefore the TSS variations with activity are proportional to the amplitude
of the convective blueshift for all spectral types. A constant attenuation factor is  compatible with the MHD simulations performed by \cite{beeck15}. 
  }
\item{We derive an amplitude of $RV$ variations due to the convective blueshift (including a possible range corresponding to different inclination assumptions) 
from the estimated convective blueshift and observed LogR'$_{\rm HK}$ variations. 
This amplitude is compared with the observed $RV$ variations for the stars of our sample. We find a global agreement in term of amplitude.  
The effects of temperature and inclination \cite[following the inclination study for the Sun of][]{borgniet15} aptly explain  the 
observed dispersion, except in the domain of small activity variability, in which some of our $RV$ 
amplitudes may be overestimated. }
\item{ \cite{dravins81} and \cite{hamilton99} showed that, for the Sun, the $RV $ depends on wavelength,
clearly visible when taking into account the $F$ dependence. 
We generalise these results and observe this dependence for the first time for a large sample of main sequence stars. 
The amplitude of the effect decreases towards smaller temperatures, and disappears for our K2 sample. It also decreases as activity increases.
 }
\end{itemize}

The results obtained in this paper will allow us to perform realistic simulations of $RV$ temporal series for various types
of stars, following the work of \cite{borgniet15} for the Sun.

\begin{acknowledgements}

This work has been funded by the Universit\'e de Grenoble Alpes project called "Alpes Grenoble Innovation Recherche (AGIR)" and by the ANR GIPSE ANR-14-CE33-0018.
This work made use of several public archives and databases: 
The HARPS data have been retrieved from the ESO archive at http://archive.eso.org/wdb/wdb/adp/phase3\_spectral/form.
This research has made use of the SIMBAD database, operated at CDS, Strasbourg, France. 
The telluric line properties were retrieved from the HINTRAN database. Exoplanets information
was retrieved from the Extrasolar Planet Encyclopaedia at http://exoplanet.eu/. The solar spectrum 
from Kurucz et al. (1984) is available on-line at  
http://kurucz.harvard.edu/sun/fluxatlas2005/. 

\end{acknowledgements}

\bibliographystyle{aa}
\bibliography{b29052}

\begin{appendix}

\section{Additional tables}


\onecolumn
\onllongtab{1}{
\begin{longtable}{llllllllll}
\caption{\label{tab_sample2} Star properties}\\
\hline
Name & Sp. t.  & B-V & Teff & N$_{\rm Spectra}$ & TSS & $\sigma$TSS & LogR'$_{\rm HK}$ & $\sigma$LogR'$_{\rm HK}$ & Conv. Blueshift \\
    &         &     &  (K) &            & (m/s/(F/Fc))&  (m/s/(F/Fc))     &        &  & (m/s) \\
\hline
   HD1388     & G0 &    0.59    &   5954    &   123    &   -846.9    &     7.9    &  -4.964    &   0.001& 500.\\
   HD3823     & G0 &    0.56    &   6022    &    64    &   -929.9    &     9.8    &  -4.976    &   0.001& 549.\\
   HD4307     & G0 &    0.61    &   5812    &    68    &   -906.5    &     8.7    &  -5.040    &   0.001& 535.\\
  HD11505     & G0 &    0.63    &   5752    &    22    &   -859.6    &    12.2    &  -4.985    &   0.001& 507.\\
  HD14374     & G0 &    0.75    &   5425    &     8    &   -345.3    &    26.9    &  -4.650    &   0.014& 204.\\
  HD20807     & G0 &    0.60    &   5866    &   176    &   -719.5    &     5.5    &  -4.876    &   0.001& 425.\\
  HD21938     & G0 &    0.55    &   5778    &    20    &   -892.1    &    17.7    &  -4.899    &   0.001& 526.\\
  HD27063     & G0 &    0.64    &   5767    &    30    &   -645.0    &    15.6    &  -4.732    &   0.004& 381.\\
  HD31527     & G0 &    0.59    &   5898    &   231    &   -858.1    &     4.7    &  -4.932    &   0.000& 506.\\
  HD31822     & G0 &    0.57    &   6042    &    44    &   -717.5    &    23.7    &  -4.847    &   0.001& 423.\\
  HD32724     & G0 &    0.61    &   5818    &    30    &   -929.2    &    12.2    &  -5.008    &   0.002& 548.\\
  HD36379     & G0 &    0.56    &   6030    &    71    &   -950.3    &    11.4    &  -4.965    &   0.001& 561.\\
  HD38973     & G0 &    0.59    &   6016    &    33    &   -861.5    &    16.8    &  -4.960    &   0.002& 508.\\
  HD39091     & G0 &    0.58    &   6003    &    83    &   -874.1    &    10.4    &  -4.970    &   0.001& 516.\\
  HD44447     & G0 &    0.53    &   5999    &    30    &   -906.3    &    13.3    &  -4.945    &   0.001& 535.\\
  HD52265     & G0 &    0.54    &   6136    &     6    &   -888.7    &    63.5    &  -4.979    &   0.007& 524.\\
  HD67458     & G0 &    0.60    &   5891    &    25    &   -748.8    &    14.0    &  -4.903    &   0.003& 442.\\
 HD68978A     & G0 &    0.61    &   5965    &   123    &   -740.5    &     9.2    &  -4.856    &   0.002& 437.\\
  HD71479     & G0 &    0.63    &   6026    &    31    &   -877.4    &    14.5    &  -5.037    &   0.003& 518.\\
  HD73524     & G0 &    0.60    &   6017    &   138    &   -841.5    &     7.3    &  -4.997    &   0.001& 496.\\
  HD78558     & G0 &    0.57    &   5711    &    31    &   -815.5    &    13.6    &  -4.929    &   0.001& 481.\\
  HD88218     & G0 &    0.60    &   5878    &    67    &   -937.5    &     8.8    &  -5.025    &   0.001& 553.\\
  HD88742     & G0 &    0.59    &   5981    &    31    &   -796.4    &    11.3    &  -4.682    &   0.009& 470.\\
  HD96700     & G0 &    0.61    &   5845    &   309    &   -889.2    &     3.8    &  -4.938    &   0.000& 525.\\
  HD97037     & G0 &    0.61    &   5883    &    87    &   -945.2    &     8.3    &  -4.988    &   0.001& 558.\\
 HD105837     & G0 &    0.52    &   5907    &    23    &   -758.0    &    20.5    &  -4.779    &   0.003& 447.\\
 HD114729     & G0 &    0.62    &   5844    &    47    &   -932.6    &    13.1    &  -5.032    &   0.001& 550.\\
 HD117618     & G0 &    0.60    &   5990    &    10    &   -807.0    &    30.7    &  -4.982    &   0.002& 476.\\
 HD134060     & G0 &    0.62    &   5966    &   280    &   -835.6    &     5.2    &  -5.001    &   0.001& 493.\\
 HD143114     & G0 &    0.61    &   5775    &    20    &   -796.3    &    16.8    &  -4.934    &   0.001& 470.\\
 HD147512     & G0 &    0.73    &   5530    &    29    &   -584.5    &     9.5    &  -4.996    &   0.005& 345.\\
 HD150433     & G0 &    0.64    &   5665    &    67    &   -733.9    &     8.1    &  -4.952    &   0.001& 433.\\
 HD183658     & G0 &    0.65    &   5803    &    43    &   -801.5    &     9.2    &  -4.984    &   0.002& 473.\\
 HD197210     & G0 &    0.70    &   5577    &    16    &   -608.4    &    14.4    &  -4.888    &   0.007& 359.\\
 HD198075     & G0 &    0.59    &   5846    &    10    &   -723.4    &    31.4    &  -4.903    &   0.005& 427.\\
 HD204385     & G0 &    0.59    &   6033    &    20    &   -902.3    &    21.4    &  -4.966    &   0.003& 532.\\
 HD206172     & G0 &    0.67    &   5608    &     2    &   -684.9    &    78.4    &  -4.856    &   0.007& 404.\\
 HD213240     & G0 &    0.61    &   5982    &     3    &  -1020.7    &    62.5    &  -5.085    &   0.005& 602.\\
 HD210752     & G0 &    0.52    &   5923    &    20    &   -874.0    &    18.8    &  -4.849    &   0.001& 516.\\
 HD215456     & G0 &    0.63    &   5789    &   181    &   -957.8    &     5.1    &  -5.070    &   0.001& 565.\\
 HD216435     & G0 &    0.62    &   6008    &    12    &   -545.3    &    54.1    &  -5.025    &   0.004& 322.\\
   HD1320     & G2 &    0.65    &   5679    &    13    &   -739.2    &    22.4    &  -4.865    &   0.004& 436.\\
   HD2071     & G2 &    0.68    &   5719    &    49    &   -708.5    &    10.9    &  -4.852    &   0.003& 418.\\
  HD20619     & G2 &    0.66    &   5703    &    34    &   -700.5    &    10.6    &  -4.793    &   0.005& 413.\\
  HD28701     & G2 &    0.61    &   5710    &    14    &   -824.7    &    23.4    &  -4.951    &   0.002& 487.\\
  HD37962     & G2 &    0.65    &   5718    &     3    &   -641.7    &    50.0    &  -4.779    &   0.014& 379.\\
  HD38858     & G2 &    0.64    &   5733    &   201    &   -758.5    &     4.3    &  -4.907    &   0.001& 448.\\
  HD45289     & G2 &    0.68    &   5717    &    71    &   -783.4    &     8.7    &  -5.032    &   0.001& 462.\\
 HD59711A     & G2 &    0.64    &   5722    &    47    &   -761.0    &    11.4    &  -4.928    &   0.002& 449.\\
  HD71334     & G2 &    0.67    &   5694    &    42    &   -759.9    &    11.8    &  -4.988    &   0.001& 448.\\
  HD78429     & G2 &    0.60    &   5760    &    58    &   -726.2    &    10.9    &  -4.902    &   0.004& 428.\\
  HD88084     & G2 &    0.64    &   5766    &    22    &   -793.2    &    15.4    &  -4.959    &   0.001& 468.\\
  HD93385     & G2 &    0.59    &   5977    &   214    &   -867.8    &     6.3    &  -4.983    &   0.001& 512.\\
  HD95521     & G2 &    0.65    &   5773    &    22    &   -714.8    &    16.2    &  -4.870    &   0.008& 422.\\
 HD102365     & G2 &    0.67    &   5629    &   257    &   -711.7    &     3.3    &  -4.935    &   0.000& 420.\\
 HD104982     & G2 &    0.65    &   5692    &    36    &   -743.2    &    12.5    &  -4.944    &   0.001& 439.\\
 HD108309     & G2 &    0.68    &   5775    &    48    &   -814.7    &     8.9    &  -5.019    &   0.002& 481.\\
 HD121504     & G2 &    0.60    &   6022    &     8    &   -717.3    &    39.9    &  -4.817    &   0.012& 423.\\
 HD125881     & G2 &    0.60    &   6036    &    66    &   -832.4    &     9.9    &  -4.889    &   0.002& 491.\\
 HD145666     & G2 &    0.56    &   5958    &    17    &   -669.7    &    24.3    &  -4.738    &   0.003& 395.\\
 HD145809     & G2 &    0.59    &   5778    &    30    &   -952.4    &    15.4    &  -5.008    &   0.001& 562.\\
 HD146233     & G2 &    0.65    &   5818    &   329    &   -751.3    &     5.8    &  -4.919    &   0.001& 443.\\
 HD168871     & G2 &    0.54    &   5983    &   116    &   -785.1    &    10.6    &  -4.936    &   0.000& 463.\\
 HD177758     & G2 &    0.56    &   5862    &    18    &   -778.2    &    24.5    &  -4.913    &   0.002& 459.\\
 HD189567     & G2 &    0.64    &   5726    &   203    &   -735.7    &     4.6    &  -4.897    &   0.001& 434.\\
 HD193193     & G2 &    0.58    &   5979    &    24    &   -885.9    &    18.9    &  -4.921    &   0.003& 523.\\
 HD196800     & G2 &    0.59    &   6010    &    19    &   -836.2    &    27.7    &  -5.006    &   0.002& 493.\\
 HD207129     & G2 &    0.60    &   5937    &   257    &   -792.0    &     5.6    &  -4.892    &   0.002& 467.\\
 HD208487     & G2 &    0.55    &   6146    &    19    &   -667.8    &    42.4    &  -4.924    &   0.003& 394.\\
 HD210918     & G2 &    0.65    &   5755    &    63    &   -797.2    &     8.5    &  -4.996    &   0.002& 470.\\
 HD213575     & G2 &    0.67    &   5671    &    24    &   -882.5    &    12.7    &  -5.063    &   0.002& 521.\\
 HD223171     & G2 &    0.66    &   5841    &    62    &   -863.5    &    10.2    &  -5.011    &   0.002& 510.\\
    HD967     & G5 &    0.61    &   5564    &    21    &   -584.4    &    17.5    &  -4.872    &   0.002& 345.\\
   HD8828     & G5 &    0.74    &   5403    &    35    &   -516.7    &    11.3    &  -4.992    &   0.001& 305.\\
   HD8859     & G5 &    0.71    &   5502    &     7    &   -577.5    &    27.5    &  -4.980    &   0.002& 341.\\
  HD12387     & G5 &    0.66    &   5700    &    20    &   -752.9    &    15.4    &  -4.979    &   0.003& 444.\\
  HD16141     & G5 &    0.71    &   5806    &     5    &   -816.8    &    25.4    &  -5.092    &   0.007& 482.\\
  HD16714     & G5 &    0.71    &   5518    &    15    &   -605.2    &    16.6    &  -4.960    &   0.003& 357.\\
  HD19034     & G5 &    0.67    &   5477    &    14    &   -595.8    &    20.0    &  -4.884    &   0.004& 352.\\
  HD20407     & G5 &    0.57    &   5866    &    27    &   -783.8    &    15.9    &  -4.876    &   0.001& 462.\\
  HD28471     & G5 &    0.65    &   5745    &    14    &   -782.1    &    20.6    &  -4.979    &   0.005& 461.\\
  HD50806     & G5 &    0.71    &   5633    &    73    &   -768.2    &     6.2    &  -5.102    &   0.001& 453.\\
  HD66428     & G5 &    0.71    &   5705    &     2    &   -755.8    &    53.9    &  -5.075    &   0.022& 446.\\
  HD78538     & G5 &    0.64    &   5786    &     3    &   -475.0    &    68.8    &  -4.577    &   0.036& 280.\\
  HD78747     & G5 &    0.51    &   5778    &    39    &   -836.6    &    18.4    &  -4.870    &   0.001& 494.\\
  HD89454     & G5 &    0.70    &   5728    &    40    &   -494.9    &    15.8    &  -4.678    &   0.004& 292.\\
  HD90156     & G5 &    0.68    &   5599    &   124    &   -652.5    &     4.6    &  -4.948    &   0.000& 385.\\
  HD96423     & G5 &    0.69    &   5711    &    63    &   -724.6    &     7.5    &  -5.037    &   0.001& 428.\\
 HD104263     & G5 &    0.74    &   5477    &    20    &   -584.5    &    19.5    &  -5.036    &   0.003& 345.\\
 HD107148     & G5 &    0.70    &   5805    &     4    &   -746.7    &    29.1    &  -5.003    &   0.002& 441.\\
 HD110619     & G5 &    0.67    &   5613    &    24    &   -640.2    &    13.3    &  -4.849    &   0.008& 378.\\
 HD111031     & G5 &    0.70    &   5801    &    38    &   -732.6    &     9.9    &  -5.074    &   0.001& 432.\\
 HD115674     & G5 &    0.68    &   5649    &    23    &   -655.1    &    13.8    &  -4.900    &   0.004& 387.\\
 HD124364     & G5 &    0.66    &   5584    &    13    &   -644.7    &    21.8    &  -4.825    &   0.010& 380.\\
 HD161098     & G5 &    0.67    &   5560    &   100    &   -627.2    &     5.6    &  -4.893    &   0.002& 370.\\
 HD189625     & G5 &    0.66    &   5846    &    17    &   -604.6    &    24.6    &  -4.814    &   0.011& 357.\\
 HD190647     & G5 &    0.76    &   5639    &    11    &   -698.9    &    22.3    &  -5.137    &   0.002& 412.\\
 HD204313     & G5 &    0.68    &   5776    &    67    &   -721.0    &     8.8    &  -5.020    &   0.002& 425.\\
 HD220507     & G5 &    0.70    &   5698    &    79    &   -740.7    &     8.9    &  -5.054    &   0.001& 437.\\
 HD222422     & G5 &    0.75    &   5475    &     6    &   -432.0    &    30.4    &  -4.783    &   0.008& 255.\\
 HD222582     & G5 &    0.65    &   5779    &    25    &   -776.1    &    16.3    &  -4.996    &   0.004& 458.\\
 HD222595     & G5 &    0.71    &   5648    &    20    &   -593.2    &    18.6    &  -4.809    &   0.013& 350.\\
 HD224393     & G5 &    0.61    &   5774    &     4    &   -822.5    &    44.4    &  -4.811    &   0.020& 485.\\
  HD10700     & G8 &    0.72    &   5310    &   438    &   -469.9    &     2.3    &  -4.942    &   0.000& 277.\\
  HD20003     & G8 &    0.77    &   5494    &    36    &   -533.1    &    13.8    &  -4.964    &   0.008& 315.\\
  HD20794     & G8 &    0.71    &   5401    &   491    &   -555.6    &     2.4    &  -4.969    &   0.000& 328.\\
  HD21411     & G8 &    0.71    &   5473    &     2    &   -493.5    &    58.9    &  -4.663    &   0.007& 291.\\
  HD37986     & G8 &    0.79    &   5507    &    28    &   -493.7    &    13.3    &  -5.083    &   0.002& 291.\\
  HD45364     & G8 &    0.76    &   5434    &    53    &   -520.0    &    10.1    &  -4.972    &   0.002& 307.\\
  HD69830     & G8 &    0.79    &   5402    &   596    &   -477.9    &     2.4    &  -4.992    &   0.001& 282.\\
  HD85119     & G8 &    0.76    &   5425    &     2    &   -254.2    &    63.0    &  -4.443    &   0.010& 150.\\
  HD94151     & G8 &    0.72    &   5583    &    22    &   -594.6    &    14.9    &  -4.971    &   0.007& 351.\\
  HD97343     & G8 &    0.78    &   5410    &    51    &   -501.9    &     9.8    &  -5.019    &   0.001& 296.\\
  HD98281     & G8 &    0.76    &   5381    &    65    &   -505.9    &     7.9    &  -4.907    &   0.005& 298.\\
 HD111232     & G8 &    0.68    &   5460    &    30    &   -582.5    &    13.1    &  -4.951    &   0.001& 344.\\
 HD123265     & G8 &    0.83    &   5338    &     7    &   -462.4    &    22.2    &  -5.087    &   0.006& 273.\\
 HD124292     & G8 &    0.74    &   5443    &    41    &   -558.8    &     8.4    &  -4.998    &   0.001& 330.\\
 HD132648     & G8 &    0.72    &   5418    &    17    &   -500.2    &    15.5    &  -4.834    &   0.011& 295.\\
 HD136894     & G8 &    0.74    &   5412    &    37    &   -524.5    &     8.7    &  -4.995    &   0.001& 309.\\
 HD145598     & G8 &    0.66    &   5417    &     7    &   -471.6    &    29.5    &  -4.899    &   0.001& 278.\\
 HD157172     & G8 &    0.78    &   5451    &    89    &   -487.2    &     6.6    &  -4.987    &   0.004& 287.\\
 HD161612     & G8 &    0.71    &   5616    &    41    &   -626.5    &     9.2    &  -5.025    &   0.001& 370.\\
 HD167359     & G8 &    0.75    &   5348    &     3    &   -433.3    &    34.5    &  -4.786    &   0.021& 256.\\
 HD172513     & G8 &    0.77    &   5500    &    33    &   -442.3    &    15.0    &  -4.779    &   0.004& 261.\\
 HD196761     & G8 &    0.72    &   5415    &    44    &   -512.8    &    10.2    &  -4.904    &   0.003& 303.\\
 HD210277     & G8 &    0.71    &   5505    &    23    &   -570.8    &    11.1    &  -5.071    &   0.003& 337.\\
 HD213628     & G8 &    0.72    &   5555    &    18    &   -556.4    &    17.8    &  -4.960    &   0.003& 328.\\
 HD213941     & G8 &    0.66    &   5532    &    30    &   -693.7    &    16.4    &  -4.907    &   0.006& 409.\\
 HD214385     & G8 &    0.63    &   5654    &    15    &   -725.8    &    24.4    &  -4.903    &   0.003& 428.\\
 HD224619     & G8 &    0.74    &   5436    &    16    &   -580.2    &    16.5    &  -4.970    &   0.003& 342.\\
    HD870     & K0 &    0.77    &   5381    &     2    &   -372.1    &    39.2    &  -4.691    &   0.008& 220.\\
   HD9796     & K0 &    0.82    &   5179    &     4    &   -402.1    &    46.6    &  -4.895    &   0.019& 237.\\
  HD26965     & K0 &    0.82    &   5153    &   395    &   -334.4    &     2.3    &  -4.936    &   0.002& 197.\\
  HD39194     & K0 &    0.77    &   5205    &   141    &   -404.8    &     5.6    &  -4.939    &   0.001& 239.\\
  HD72579     & K0 &    0.77    &   5449    &    16    &   -509.2    &    19.4    &  -5.084    &   0.002& 300.\\
  HD74014     & K0 &    0.76    &   5561    &    22    &   -551.4    &    11.5    &  -5.070    &   0.001& 325.\\
  HD80883     & K0 &    0.83    &   5233    &     3    &   -319.3    &    46.6    &  -4.652    &   0.020& 188.\\
  HD83443     & K0 &    0.79    &   5511    &    15    &   -408.8    &    19.2    &  -4.933    &   0.014& 241.\\
  HD90711     & K0 &    0.81    &   5444    &    23    &   -491.2    &    12.7    &  -5.019    &   0.008& 290.\\
  HD90812     & K0 &    0.82    &   5164    &     5    &   -364.5    &    22.4    &  -4.956    &   0.007& 215.\\
 HD104006     & K0 &    0.83    &   5023    &     6    &   -277.2    &    32.0    &  -4.970    &   0.003& 164.\\
 HD130322     & K0 &    0.78    &   5365    &     3    &   -399.0    &    28.2    &  -4.700    &   0.001& 235.\\
 HD176157     & K0 &    0.83    &   5181    &     2    &   -440.2    &    44.8    &  -4.769    &   0.008& 260.\\
 HD202605     & K0 &    0.72    &   5658    &     2    &   -257.9    &    38.0    &  -4.523    &   0.001& 152.\\
 HD203384     & K0 &    0.78    &   5586    &    13    &   -533.1    &    19.2    &  -5.047    &   0.006& 315.\\
 HD220256     & K0 &    0.85    &   5144    &     8    &   -368.6    &    20.6    &  -5.010    &   0.003& 217.\\
  HD17970     & K2 &    0.84    &   5040    &    14    &   -304.0    &    14.3    &  -4.994    &   0.006& 179.\\
  HD40307     & K2 &    0.95    &   4977    &   337    &   -272.9    &     3.4    &  -4.951    &   0.003& 161.\\
  HD82516     & K2 &    0.91    &   5104    &    24    &   -202.1    &    17.9    &  -4.932    &   0.010& 119.\\
 HD101930     & K2 &    0.91    &   5164    &    21    &   -321.6    &    13.0    &  -4.999    &   0.010& 190.\\
 HD106275     & K2 &    0.90    &   5059    &     6    &   -270.8    &    37.2    &  -4.874    &   0.035& 160.\\
 HD129642     & K2 &    0.95    &   5026    &    31    &   -203.3    &    12.5    &  -4.959    &   0.003& 120.\\
 HD130930     & K2 &    0.94    &   5027    &     4    &   -263.2    &    29.5    &  -5.025    &   0.013& 155.\\
 HD154577     & K2 &    0.89    &   4900    &   294    &   -283.1    &     3.9    &  -4.869    &   0.002& 167.\\
 HD176986     & K2 &    0.94    &   5018    &    29    &   -299.8    &    15.2    &  -4.823    &   0.005& 177.\\
 HD192031     & K2 &    0.72    &   5215    &     2    &   -312.2    &    33.0    &  -4.924    &   0.005& 184.\\
 HD192310     & K2 &    0.91    &   5166    &  1230    &   -251.0    &     1.6    &  -4.993    &   0.001& 148.\\
 HD209742     & K2 &    0.85    &   5137    &     7    &   -232.7    &    27.5    &  -4.834    &   0.021& 137.\\
 HD204941     & K2 &    0.91    &   5056    &    16    &   -255.7    &    20.4    &  -4.967    &   0.007& 151.\\
  HD44573     & K2 &    0.92    &   5071    &    14    &   -177.7    &    19.6    &  -4.562    &   0.005& 105.\\
  HD68607     & K2 &    0.83    &   5215    &     7    &   -143.7    &    25.6    &  -4.681    &   0.013&  85.\\
  HD23356     & K2 &    0.93    &   5004    &    10    &   -217.2    &    14.3    &  -4.738    &   0.004& 128.\\
 HD148303     & K2 &    0.97    &   4958    &     8    &   -175.6    &    28.4    &  -4.648    &   0.022& 104.\\
 HD220339     & K2 &    0.90    &   5029    &    11    &   -300.6    &    18.5    &  -4.808    &   0.017& 177.\\
  HD13808     & K2 &    0.85    &   5087    &    66    &   -286.1    &     8.0    &  -4.872    &   0.010& 169.\\
  HD65562     & K2 &    0.86    &   5076    &     3    &   -270.1    &    41.0    &  -4.893    &   0.022& 159.\\
 HD203850     & K2 &    0.92    &   4879    &     2    &   -297.4    &    41.0    &  -4.787    &   0.008& 175.\\
\end{longtable}
\tablefoot{Star name, spectral type and B-V from the CDS, Teff from \cite{sousa08}, number of spectra used in the analysis, 
TSS and its 1-$\sigma$ uncertainty, averaged LogR'$_{\rm HK}$ and its 1-$\sigma$ uncertainty and the convective blueshift derived from the TSS in 
Sect.~4.1 (without the wavelength correction). 
}
}

\section{Continuum correction: procedure}

The procedure is as follows:

\begin{itemize}
\item{
We define an upper envelope for each spectrum by retrieving the highest intensity in each 5 $\AA$ bin. 
We eliminate outliers that may be due to cosmics by performing a linear fit on eleven such adjacent  points and 
removing the point with the largest residual. For each remaining point, we recompute the 
linear fit and consider the residual after the fit removal at that point. 
The distribution of these values is fitted with a gaussian. Points with values outside a 3-$\sigma$ range or too close to each other 
are also eliminated. The resulting intensities versus wavelength are then smoothed and interpolated on the original list of wavelengths, 
providing an upper envelope for the spectrum.
}
\item{
The spectrum is then divided by this upper envelope. The distribution of all intensities in the spectrum peaks at the level of the continuum 
(its position depends on the noise level). This provides a 
correcting factor, by which the spectra is multiplied to provide the final spectra, normalised to a continuum of 1. 
A visual examination shows that the continuum can be slightly off in  a few wavelength ranges. The effect is larger for low S/N spectra. 
}
\end{itemize}

Note that in principle the correcting factor depends on the wavelength, as the S/N depends on the order. 
However the impact on our analysis is very limited, especially 
since we have considered lines in the 5000-6855~$\AA$ range only.

\onecolumn
\onllongtab{1}{
\begin{longtable}{lll}
\caption{\label{tab_line} Sample lines}\\
\hline
Wavelength & Element  & Nb. sample  \\
($\AA$) & & \\ \hline
4999.5000 &  TiI &  2\\
5001.8636 &  FeI &  6\\
5002.7927 &  FeI &  6\\
5005.7123 &  FeI &  6\\
5014.9425 &  FeI &  5\\
5016.1600 &  TiI &  6\\
5016.4790 &  FeI &  4\\
5022.2355 &  FeI &  6\\
5022.8700 &  TiI &  6\\
5024.8400 &  TiI &  6\\
5027.7567 &  FeI &  6\\
5028.1264 &  FeI &  6\\
5030.7786 &  FeI &  2\\
5036.4600 &  TiI &  6\\
5038.4000 &  TiI &  6\\
5039.9600 &  TiI &  6\\
5040.6100 &  TiI &  2\\
5043.5800 &  TiI &  3\\
5044.2114 &  FeI &  6\\
5045.4100 &  TiI &  2\\
5048.4361 &  FeI &  6\\
5049.8198 &  FeI &  4\\
5062.1000 &  TiI &  2\\
5064.6500 &  TiI &  6\\
5067.1496 &  FeI &  6\\
5068.7658 &  FeI &  6\\
5072.0784 &  FeI &  6\\
5072.6721 &  FeI &  6\\
5074.7483 &  FeI &  6\\
5078.9748 &  FeI &  6\\
5079.2230 &  FeI &  6\\
5079.7400 &  FeI &  6\\
5083.3386 &  FeI &  6\\
5085.3300 &  TiI &  1\\
5090.7740 &  FeI &  6\\
5107.4474 &  FeI &  5\\
5109.6520 &  FeI &  6\\
5125.8345 &  FeI &  1\\
5127.3593 &  FeI &  6\\
5131.4687 &  FeI &  6\\
5133.6885 &  FeI &  4\\
5137.3822 &  FeI &  6\\
5141.7390 &  FeI &  6\\
5151.9109 &  FeI &  6\\
5152.1800 &  TiI &  6\\
5159.0576 &  FeI &  6\\
5162.2729 &  FeI &  6\\
5165.4107 &  FeI &  5\\
5173.7400 &  TiI &  5\\
5187.9142 &  FeI &  6\\
5191.4550 &  FeI &  4\\
5192.9700 &  TiI &  6\\
5194.0300 &  TiI &  1\\
5194.9418 &  FeI &  6\\
5195.4723 &  FeI &  6\\
5196.0596 &  FeI &  6\\
5198.7111 &  FeI &  6\\
5215.1806 &  FeI &  6\\
5216.2740 &  FeI &  6\\
5217.3893 &  FeI &  6\\
5223.6200 &  TiI &  1\\
5224.5400 &  TiI &  4\\
5225.5261 &  FeI &  6\\
5226.8623 &  FeI &  3\\
5228.3767 &  FeI &  6\\
5234.6200 & FeII &  6\\
5242.4911 &  FeI &  6\\
5246.5500 &  TiI &  1\\
5247.0504 &  FeI &  6\\
5250.2089 &  FeI &  6\\
5250.6460 &  FeI &  6\\
5253.4617 &  FeI &  6\\
5263.3063 &  FeI &  6\\
5264.8000 & FeII &  5\\
5266.5554 &  FeI &  1\\
5273.1636 &  FeI &  6\\
5273.3736 &  FeI &  6\\
5364.8713 &  FeI &  6\\
5365.3991 &  FeI &  6\\
5367.4668 &  FeI &  6\\
5373.7086 &  FeI &  6\\
5383.3692 &  FeI &  3\\
5389.4792 &  FeI &  6\\
5393.1676 &  FeI &  4\\
5404.1516 &  FeI &  1\\
5410.9098 &  FeI &  6\\
5415.1993 &  FeI &  4\\
5426.2500 &  TiI &  1\\
5434.5238 &  FeI &  4\\
5445.0424 &  FeI &  6\\
5446.5829 &  TiI &  4\\
5446.9168 &  FeI &  1\\
5453.6400 &  TiI &  1\\
5460.5000 &  TiI &  2\\
5462.9595 &  FeI &  6\\
5466.3962 &  FeI &  6\\
5473.9005 &  FeI &  6\\
5474.2200 &  TiI &  1\\
5476.5642 &  FeI &  6\\
5481.8600 &  TiI &  1\\
5487.7460 &  FeI &  6\\
5501.4653 &  FeI &  6\\
5503.9000 &  TiI &  1\\
5506.7791 &  FeI &  6\\
5511.7800 &  TiI &  1\\
5525.5443 &  FeI &  6\\
5534.8400 & FeII &  5\\
5560.2116 &  FeI &  6\\
5562.7065 &  FeI &  6\\
5565.7040 &  FeI &  6\\
5569.6181 &  FeI &  4\\
5576.0888 &  FeI &  6\\
5624.5422 &  FeI &  4\\
5633.9465 &  FeI &  6\\
5638.2621 &  FeI &  6\\
5648.5600 &  TiI &  1\\
5655.4900 &  FeI &  6\\
5662.9380 &  TiI &  6\\
5679.0229 &  FeI &  6\\
5686.5302 &  FeI &  6\\
5701.5446 &  FeI &  6\\
5715.0986 &  TiI &  6\\
5739.4700 &  TiI &  1\\
5752.0320 &  FeI &  6\\
5775.0806 &  FeI &  6\\
5785.9800 &  TiI &  1\\
5804.2600 &  TiI &  1\\
5862.3565 &  FeI &  6\\
5880.2700 &  TiI &  1\\
5903.3100 &  TiI &  1\\
5905.6720 &  FeI &  6\\
5916.2474 &  FeI &  6\\
5918.5300 &  TiI &  3\\
5922.1100 &  TiI &  3\\
5927.7891 &  FeI &  4\\
5930.1799 &  FeI &  6\\
5937.8100 &  TiI &  1\\
5941.7500 &  TiI &  3\\
5952.7184 &  FeI &  6\\
5956.6944 &  FeI &  6\\
5976.7771 &  FeI &  6\\
5983.6810 &  FeI &  6\\
6008.5566 &  FeI &  6\\
6027.0509 &  FeI &  6\\
6056.0047 &  FeI &  6\\
6064.6200 &  TiI &  1\\
6065.4822 &  FeI &  6\\
6078.4911 &  FeI &  6\\
6091.1700 &  TiI &  1\\
6102.1777 &  FeI &  6\\
6127.9066 &  FeI &  6\\
6136.6153 &  FeI &  5\\
6136.9947 &  FeI &  6\\
6157.7284 &  FeI &  6\\
6173.3356 &  FeI &  6\\
6191.5584 &  FeI &  6\\
6213.4303 &  FeI &  6\\
6219.2810 &  FeI &  6\\
6226.7363 &  FeI &  1\\
6229.2283 &  FeI &  4\\
6230.7230 &  FeI &  4\\
6232.6412 &  FeI &  6\\
6240.6462 &  FeI &  6\\
6246.3188 &  FeI &  6\\
6247.5600 & FeII &  2\\
6252.5554 &  FeI &  6\\
6256.3615 &  FeI &  6\\
6258.1000 &  TiI &  6\\
6265.1340 &  FeI &  6\\
6270.2250 &  FeI &  6\\
6290.9656 &  FeI &  6\\
6293.9257 &  FeI &  2\\
6302.4936 &  FeI &  6\\
6303.7500 &  TiI &  1\\
6312.2300 &  TiI &  1\\
6318.0175 &  TiI &  6\\
6322.6855 &  FeI &  6\\
6335.3308 &  FeI &  6\\
6380.7433 &  FeI &  6\\
6392.5388 &  FeI &  1\\
6393.6013 &  FeI &  4\\
6408.0184 &  FeI &  6\\
6419.9496 &  FeI &  6\\
6421.3508 &  FeI &  6\\
6430.8464 &  FeI &  6\\
6494.9805 &  FeI &  2\\
6546.2395 &  TiI &  6\\
6569.2155 &  FeI &  1\\
6581.2101 &  FeI &  1\\
6592.9100 &  TiI &  4\\
6592.9138 &  TiI &  6\\
6599.1000 &  TiI &  1\\
6625.0220 &  FeI &  1\\
6633.7497 &  FeI &  6\\
6663.4421 &  FeI &  6\\
6743.1200 &  TiI &  2\\
\end{longtable}
\tablefoot{TiI and FeII wavelengths from \cite{dravins08} and FeI wavelengths from \cite{nave94}, elements and number of spectral type 
samples in which the line has been used in our analysis. 
}
}

\section{Impact of inclination on the $RV$ variability}

\begin{figure} 
\includegraphics{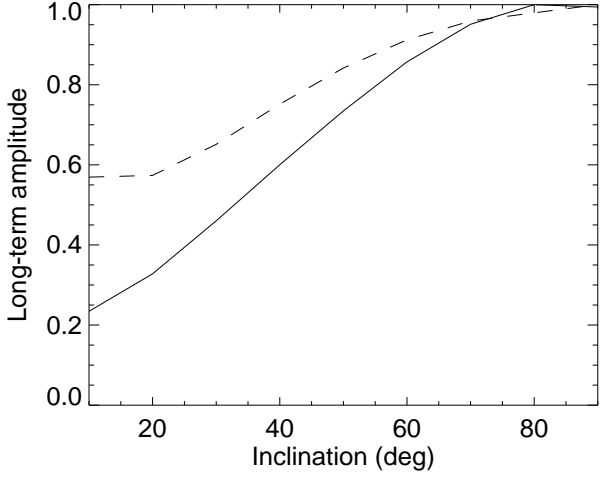}
\caption{
Dependence of the attenuation of the long-term convective blueshift 
$\Delta RV_{\rm conv}$ 
versus inclination $f_{\rm RV}$ (solid line), normalised to one, for a solar
simulation (from Borgniet et al.2015), 
and variation of the corresponding plage coverage $f_{\rm ff}$ (dashed line), also normalised.
}
\label{inc}
\end{figure}

\begin{figure} 
\includegraphics[width=15cm]{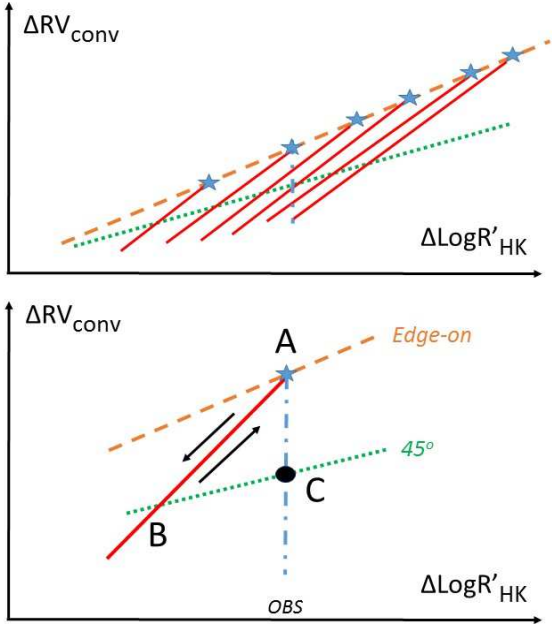}
\caption{
{\it Upper panel}: Schematic view of $\Delta RV_{\rm conv}$ versus $\Delta LogR'_{\rm HK obs}$ showing  
the impact of inclination for a star at a given temperature and different activity levels: seen equator-on 
(orange dashed line with blue stars), at 45$^{\circ}$ (dotted green line). The red tracks correspond to the various inclinations for each blue star. 
The vertical dotted-dashed blue thick line corresponds to the various $\Delta RV_{\rm conv}$ that would correspond to different inclinations for a given 
observed $\Delta LogR'_{\rm HK obs}$. 
{\it Lower panel}: Same zoomed in on one particular observation (see text). 
}
\label{illust_inc}
\end{figure}

We now consider the dependence of the rotational axis on inclination.
\cite{borgniet15} have studied the impact of inclination on the $RV$ time series from a solar simulation for various inclinations. 
The $RV$ and plage coverage (which is directly related to the chromospheric emission) are maximal for a Sun seen equator-on and minimal for a 
pole-on configuration. They do not vary by the same factor however: a factor of 4.2 for $RV$ and a factor of 1.8 for the plage coverage. 
In the following we will apply this $RV$ factor $f_{\rm RV}$ to our computed $RV$ to derive a reconstructed $RV$ at a given inclination to 
another. We apply the plage coverage factor $f_{\rm ff}$ to LogR'$_{\rm HK}$ as well.  
Normalised $f_{\rm RV}$ and $f_{\rm ff}$ are shown on Fig.~\ref{inc}.

We now estimate the impact of this dependence on our reconstructed $\Delta RV_{\rm conv}$. Let us first consider a star, with a given Teff, 
seen equator-on (as the Sun). 
Depending on the amplitude of activity variability of this star, it will be located at different positions in the ($\Delta LogR'_{\rm HK}$, 
$\Delta RV_{\rm conv}$) diagram, and will follow the orange line in Fig.~\ref{illust_inc} (upper panel), different blue stars corresponding to specific $\Delta LogR'_{\rm HK}$. 
If the same stars were seen with an inclination 
of 45$^{\circ}$, they would be located on the black line, whose slope is defined by $G$ (defined in Sect.~5.2.1, for the Teff of the star). 
One of these stars with a given activity level but seen at different inclinations would follow one of the red tracks. 
When we reconstruct $\Delta RV_{\rm conv}$ as above, we do not know the inclination of the star, only
the amplitude of variation of the LogR'$_{\rm HK}$. Given an observed $\Delta LogR'_{\rm HK obs}$, 
the reconstructed $\Delta RV_{\rm conv}$ for various inclinations should therefore be at any location along the vertical thick blue line 
(close to the top if seen edge on, and close to the bottom if seen pole-on), that is, where the vertical line crosses any red track.

We therefore wish to compute the values corresponding to this blue segment for each of our stars, to estimate the value it would take 
for all possible inclinations. Note that our computations are made for inclinations between 10 and 90$^{\circ}$; the range covered by the simulation of 
\cite{borgniet15}. On Fig.~\ref{illust_inc} (lower panel), the black dot at C is the reconstructed $\Delta RV_{\rm conv}$ from eq. 5 
applied to the observed $\Delta LogR'_{\rm HK obs}$. As an example, let us consider how to reconstruct $\Delta RV_{\rm conv}$ for an 
equator-on configuration, that is, for point A, 
which has the same observed $\Delta LogR'_{\rm HK obs}$. This star is then on the red track: if it was seen 
at a different inclination (i.e. from another observer's point of  view) it would be along this track. Such an observer observing 
it at 45$^{\circ}$  would therefore see point B, which has the following coordinates:

\begin{equation}
\Delta LogR'_{\rm HK}(B)= \Delta LogR'_{\rm HK obs} \times f_{\rm ff}(45^{\circ})/f_{\rm ff}(90^{\circ})
.\end{equation}

%

$\Delta RV_{\rm conv}(B)$ is derived from eq. 5 applied to $\Delta LogR'_{\rm HK}(B)$.
From this point B, we can move along the track back to point A by applying the $RV$ correcting factor to the point B $RV$ amplitude:

\begin{equation}
\Delta RV_{\rm conv}(A)=\Delta RV_{\rm conv}(B) \times  f_{\rm RV}(90^{\circ})/f_{\rm RV}(45^{\circ})
\end{equation}

The same computation, using eqs. C.1 and C.2 but applied to any inclination instead of 90$^{\circ}$ will provide the range of $\Delta RV_{\rm conv}$
covering all inclinations between 10 and 90$^{\circ}$.

\end{appendix}

\end{document}